\documentclass[graybox,nosecnum,sort&compress]{svmult}

\usepackage{mathptmx}       % selects Times Roman as basic font
\usepackage{helvet}         % selects Helvetica as sans-serif font
\usepackage{courier}        % selects Courier as typewriter font
\usepackage{type1cm}        % activate if the above 3 fonts are
\usepackage{makeidx}         % allows index generation
\usepackage{graphicx}        % standard LaTeX graphics tool
\usepackage{multicol}        % used for the two-column index
\usepackage[bottom]{footmisc}% places footnotes at page bottom
\usepackage{hyperref}        %for hyperlinks
\usepackage{soul}            % for high-lighting of text
\hypersetup{colorlinks=true,urlcolor=blue}
\usepackage[square,numbers]{natbib}
\usepackage{amsfonts,amssymb}
\usepackage{pifont} %for ding symbols
\newcommand{\au}[2]{#2 #1}
\newcommand{\book}[5]{\emph{#1} (#2, #3, #4, #5)}
\newcommand{\books}[4]{\emph{#1} (#2, #3, #4)}
\newcommand{\oarX}[1]{\href{http://arxiv.org/abs/#1}{{\ttfamily\cob arXiv:#1}}}
\newcommand{\arX}[1]{\href{http://arxiv.org/abs/#1}{{\ttfamily\cob arXiv:#1}}}
\newcommand{\doin}[6]{\href{http://dx.doi.org/#1}{{\cob #2 #3 {\bf #4}, #5 (#6)}}}
\newcommand{\doinn}[5]{\href{http://dx.doi.org/#1}{{\cob #2 {\bf #3}, #4 (#5)}}}
\newcommand{\doij}[5]{\href{http://dx.doi.org/#1}{{\cob #2 #3 (#5) #4}}}

\newcommand{\proc}[5]{in \emph{#1}, ed.\ by #2 (#3, #4, #5)}
\newcommand{\tia}[1]{#1,}
\newcommand{\be}{\begin{equation}}
\newcommand{\ee}{\end{equation}}
\newcommand{\ba}{\begin{eqnarray}}
\newcommand{\ea}{\end{eqnarray}}
\def\bs{\begin{subequations}}
\def\es{\end{subequations}}
\def\a{\alpha}
\def\b{\beta}
\def\de{\delta}
\def\De{\Delta}
\def\g{\gamma}
\def\G{\Gamma}
\def\la{\lambda}
\def\k{\kappa}
\def\e{\epsilon}

\def\Om{\Omega}
\def\om{\omega}
\def\G{\Gamma}
\def\t{\tau}  
\def\s{\sigma}
\def\vr{\varrho}
\def\vp{\varphi}
\def\N{\nabla}
\def\H{{\rm H}}
\def\cA{\mathcal{A}}

\def\cF{\mathcal{F}}

\def\cK{\mathcal{K}}

\def\cN{\mathcal{N}}

\def\cP{\mathcal{P}}

\def\cS{\mathcal{S}}
\def\cT{\mathcal{T}}

\def\ds{d_{\rm S}}
\def\dh{d_{\rm H}}

\def\p{\partial}

\def\B{\Box}

\newcommand{\Eq}[1]{(\ref{#1})}

\def\cob{\color{blue}}
\def\Pl{{\rm Pl}}
\def\lp{\ell_\Pl}

\def\Mpl{M_\Pl}

\def\rme{{\rm e}}
\def\rmd{{\rm d}}

\def\leq{\leqslant}

\makeindex             % used for the subject index
\begin{document}
%\tableofcontents{}
\title*{Quantum Gravity and Gravitational-Wave Astronomy}
\author{Gianluca Calcagni \thanks{corresponding author}}
\institute{Gianluca Calcagni \at Instituto de Estructura de la Materia, CSIC, Serrano 121, 28006 Madrid, Spain, \email{g.calcagni@csic.es}}
\maketitle
\abstract{We review the present status of quantum-gravity phenomenology in relation to gravitational waves (GWs). The topic can be approached from two directions, a model-dependent one and a model-independent one. We introduce some among the most prominent cosmological models embedded in or motivated by theories of quantum gravity, while also pointing out certain common features shared by these and other models of quantum gravity. Three cosmological GW observables can be affected by perturbative as well as non-perturbative quantum-gravity effects: the amplitude of the stochastic GW background, the propagation speed of GWs and the luminosity distance of GW sources. While many quantum-gravity models do not give rise to any observable signal, some predict a blue-tilted stochastic background or a modified luminosity distance, both detectable by future GW interferometers. We conclude that it is difficult, but still possible, to test quantum gravity with GW observations.}

\section{Keywords} 
Quantum gravity; theories beyond Einstein gravity; phenomenology; modified dispersion relations; inflation; stochastic GW background; GW propagation speed; luminosity distance.

%%%%%%%%%%%%%%%%%%%%%%%%%%%%%%%%%%%%%%%%%%%%%%%%%%%%%%%%%%%%%%%%%%%%%%%%%%%%%%%%%
%%%%%%%%%%%%%%%%%%%%%%%%%%%%%%%%%%%%%%%%%%%%%%%%%%%%%%%%%%%%%%%%%%%%%%%%%%%%%%%%%

\section{Introduction: Why quantum gravity?}

Gravitational-wave (GW) astronomy is the new frontier of knowledge on astrophysics, the fundamental gravitational interaction and the properties of spacetime. In the recent history of gravitational physics, seldom have experiment and theory converged in such a synergy. So far, observations of mergers by the LIGO-Virgo network have confirmed all predictions of general relativity (GR) about GWs emitted by binary systems of black holes and other compact objects \cite{TheLIGOScientific:2016src,LIGOScientific:2020stg}, but there is still much to be explored about the nature of gravity. However, while GWs are acknowledged as a promising test of modifications of GR, their role in exploring such an extreme regime as quantum gravity is unclear \cite{Ame98,NgDa2,Ame13,EMNan,revmu,ArCa2,Yunes:2016jcc,Kobakhidze:2016cqh,ACCR,BYY,TaYa,Mas18,Bosso:2018ckz,AMY,Giddings:2019ujs,Calcagni:2019kzo,Belgacem:2019pkk,Calcagni:2019ngc,Wang:2020pgu,Garcia-Chung:2020zyq}. The answer to this question can have lasting repercussions, since the discovery of a signature of quantum gravity would change forever our way of understanding the fundamental interactions of Nature. The general expectation is that the low curvature, low energies and large distances characterizing the production and propagation of GWs make it unlikely, or at least difficult, to probe Planck-size effects, unless they are cumulatively amplified by some cosmological mechanism. As we will see, in some isolated cases such a mechanism is in action.

Before entering into the topic of this chapter, let us briefly recall what quantum gravity is and why we are interested in it. Quantum gravity is a generic label denoting any theory unifying the gravitational force and quantum mechanics in a consistent way. Although there is no experimental evidence that gravity should be a quantum interaction, theoretically one would expect all forces of Nature to follow about the same rules, while at the moment gravity is described with quite different tools than those employed in the Standard Model of electroweak and strong interactions. Sometimes, \emph{Gedankenexperimente} are invoked \cite{For82,Wal84} to explain local observations \cite{PaG81} and show that it is not possible to have a purely classical gravitational force interact with quantum matter fields as in the semi-classical Einstein equations $G_{\mu\nu}=\k^2 \langle T_{\mu\nu}\rangle$, where $\k^2=8\pi G$ is proportional to Newton's constant $G$, $G_{\mu\nu}$ is the Einstein tensor and $\langle T_{\mu\nu}\rangle$ is the expectation value of the energy-momentum tensor. However, one can change the quantum setting in such a way as to make classical gravity compatible with present observations \cite{Car08}. Other circumstantial arguments against classical gravity are the presence of singularities (black-hole singularities and the big bang) where the laws of GR break down and the inability of the latter to explain satisfactorily the cosmological constant.

Therefore, we have no compelling proof that quantum gravity is needed: there is no experimental or strong theoretical reason indicating that quantum gravity is an absolute necessity. Quantum gravity is not even necessary for the sake of getting new phenomenology, since there are many \emph{ad hoc} models doing that. Cosmological models beyond Einstein gravity such as $f(R)$, Horndeski, DHOST or infrared non-local models are not embedded in a fundamental gravitational quantum theory and, yet, deserve interest for their consequences in GW astronomy \cite{Yunes:2016jcc,BYY,Belgacem:2019pkk}. It is also worth mentioning that other types of deviations from the standard cosmological model, for instance in the matter sector while leaving GR untouched, can also trigger a signal, for instance as a large high-frequency amplitude of the stochastic GW background \cite{Bartolo:2016ami,Kuroyanagi:2018csn}.

Still, for many it is more rewarding to extract phenomenology and predictions from a robust, consistent theoretical setting rather than from \emph{ad hoc} scenarios which are either unfalsifiable or, if falsifiable or validated by observations, difficult to place inside a bigger picture. This research trend is further strengthened by the fact that quantum gravity is no longer a mirage and some concrete proposals exist \cite{Ori09,Fousp,CQC}, such as string theory, asymptotic safety, loop quantum gravity, group field theory and non-local quantum gravity, just to name some. Due to lack of space we cannot review all theories of quantum gravity but we present only scenarios where GW observables have been calculated.
\begin{itemize}
\item {\bf Stelle gravity} \cite{Ste77,Ste78,ALS,AAM}, a non-unitary, renormalizable quantum field theory of gravity on a continuous spacetime where the fundamental action contains second-order curvature invariants ($R^2$, $R_{\mu\nu}R^{\mu\nu}$, $R_{\mu\nu\s\t}R^{\mu\nu\s\t}$).
\item {\bf String theory} in its low-energy limit \cite{Pol98,BBSb,Zwi09} and the corresponding cosmological models \cite{Baumann:2014nda}. Here gravity is not quantized directly but it emerges in the spectrum of quantized stringy fundamental blocks and is unified with the other Standard-Model interactions.
\item {\bf Asymptotic safety} \cite{Wei79,Reu1,NiR,Nie06,CPR,Lit11,RSnax}, a non-perturbative quantization of gravity via the functional renormalization-group approach.
\item Quantum gravities with discrete pre-geometries ({\bf LQG/SF/GFT}), where spacetime emerges from a structure (``pre-geometry'') characterized by discrete labels (spin group representations). The two main theories under this paradigm are loop quantum gravity (LQG) \cite{rov07,thi01} and spin foams (SF) \cite{Per03,Rov10,Per13}, possibly different manifestations of a more general setting known as group field theory (GFT) \cite{Fre05,Ori09,BaO11,Fousp,Ori13,GiSi}.
\item {\bf Causal dynamical triangulations} \cite{AmJ,AJL4,AJL5,lol08,AJL8,AGJL4,CoJu,CoDo}, a non-perturbative quantization of gravity via  a path integral on discrete triangulated geometries.
\item {\bf Non-local quantum gravity} \cite{Kuz89,Tom97,Modesto:2011kw,BGKM,Modesto:2017sdr,BrCM}, a perturbative quantum field theory of gravity on a continuous spacetime with scale-dependent non-local operators in the fundamental action.
\item {\bf Ho\v{r}ava--Lifshitz gravity} \cite{Hor09,Hor3,HoMe}, a perturbative quantum field theory of gravity on a continuous spacetime where the time dimension scales anomalously.
\end{itemize}
Stelle gravity and Ho\v{r}ava--Lifshitz gravity are known to have issues about, respectively, unitarity and violation of Lorentz invariance, but we include them nonetheless because some of their features are common to the other approaches and can be calculated easily. To this list, we add other approaches that do not have the same scope or do not reach the same level of completeness, but that have given valuable insights into the problem of quantum gravity.
\begin{itemize}
\item {\bf Canonical quantum cosmology} (reviewed in, e.g., \cite{CQC}), models of the early universe built on the Hamiltonian formalism. They include the original approach based on the Wheeler--DeWitt equation as well as loop quantum cosmology, based on the LQG quantization of gravity in Ashtekar--Barbero variables.
\item {\bf Pre-big-bang cosmology} \cite{Gasperini:1992em}, where the duality symmetries of string theory suggest that the cosmological history of expansion and the big bang itself be preceded by a phase of growing curvature.
\item {\bf String-gas cosmology} \cite{Bra11,Brandenberger:2015kga}, a model producing primordial spectra via a thermal mechanism alternative to inflation and involving strings. 
\item {\bf New ekpyrotic scenario} \cite{Brandenberger:2020tcr,Brandenberger:2020eyf}. At the density of the string scale, new degrees of freedom govern the effective four-dimensional cosmological dynamics. At the time when such density is reached, the dynamics is dominated by an S-brane, a space-like hypersurface with zero energy density and negative pressure that induces a transition between a cosmological contracting phase (ekpyrosis) and an expanding one. 
\item {\bf Brandenberger--Ho non-commutative inflation} \cite{BH,Calcagni:2013lya}, where time and space coordinates do not commute and, as a consequence, the inflaton scalar field driving the early phase of acceleration obeys a modified dynamics.
\item {\bf Non-commutative $\k$-Minkowski spacetime} \cite{Sza01,ADKLW,Ben08,ArTr,Eckstein:2020gjd}, a spacetime with non-commuting coordinates whereupon one can construct a field theory of gravity and matter.
\item {\bf Padmanabhan's non-local field theory} \cite{Pad98,Pad99,ArCa1}, an effective field theory with non-local operators assumed to be valid near the horizon of black holes.
\item {\bf Multi-fractional spacetimes} \cite{revmu}, a class of models with scale-dependent spacetime geometries. Integrals and derivatives acquire an anomalous multi-scaling, the action gets a new discrete symmetry at short scales and standard cosmology is modified accordingly 
\end{itemize}

The background upon which we will construct cosmological observables is the flat homogeneous Friedmann--Lema\^itre--Robertson--Walker (FLRW) line element
\be\label{flrw}
\rmd s^2=-\rmd t^2+a^2(t)\,\de_{ij}\rmd x^i \rmd x^j=a^2(\tau)\,(-\rmd \tau^2+\de_{ij}\rmd x^i \rmd x^j)\,,
\ee
where $t$ is proper time, $a$ is called scale factor, $\tau:=\int\rmd t/a$ is conformal time and $i,j=1,2,3$. We set the speed of light to $c=1$. For each observable, we will recall the GR expression, its generalization to quantum gravity and the constraints placed to date in the literature.

%%%%%%%%%%%%%%%%%%%%%%%%%%%%%%%%%%%%%%%%%%%%%%%%%%%%%%%%%%%%%%%%%%%%%%%%%%%%%%%%%
%%%%%%%%%%%%%%%%%%%%%%%%%%%%%%%%%%%%%%%%%%%%%%%%%%%%%%%%%%%%%%%%%%%%%%%%%%%%%%%%%

\section{Stochastic GW background}

%%%%%%%%%%%%%%%%%%%%%%%%%%%%%%%%%%%%%%%%%%%%%%%%%%%%%%%%%%%%%%%%%%%%%%%%%%%%%%%%%

\subsection{Basics}

Let $\cP_{\rm t}(f)$ be the primordial tensor spectrum of tensor perturbations of the metric with proper wave-length $\la$, where $f=k/(2\pi)=a/\la$ is the frequency measured by a comoving observer and $k=|\mathbf{k}|$ is called comoving wave-number. These fluctuations arise in the early universe during an era of inflation or from alternative mechanisms. The amplitude of the tensor spectrum is small compared to that of scalar fluctuations $\cP_{\rm s}(f)$ and, in fact, the bound on the tensor-to-scalar ratio $r:=\cP_{\rm t}/\cP_{\rm s}$ by the \textsc{Planck} Legacy release \cite{Akrami:2018odb} from the \textsc{Planck}+TT+TE+EE+lowE +lensing+BK15+BAO data set at the pivot scale $f_0=7.7\times 10^{-17}\,{\rm Hz}$ (comoving wave-number $k_0=0.05\,{\rm Mpc}^{-1}$) is $r<0.068$ at the 95\% confidence level (CL). Other observables of interest are the tensor spectral index $n_{\rm t} := \rmd\ln\cP_{\rm t}/\rmd\ln f$ and its running $\a_{\rm t}:= \rmd n_{\rm t}/\rmd\ln f$. The primordial tensor spectrum is thus parametrized as 
\be\label{ptk}
\cP_{\rm t}(f)=\cP_{\rm t}(f_0)\,\left(\frac{f}{f_0}\right)^{n_{\rm t}(f_0)+\frac12\a_{\rm t}(f_0)\ln\frac{f}{f_0}}. %=\cP_{\rm t}(f_0)\,(f/f_0)^{n_{\rm t}(f_0)+\a_{\rm t}(f_0)\ln(f/f_0)/2}$. 
\ee
In standard inflation in GR, both $n_{\rm t}$ and $\a_{\rm t}$ are negative and the spectrum is said to be red-tilted. The amplitude of the stochastic GW background as observed today is characterized by the dimensionless density parameter 
\be\label{Omgw}
\Om_\textsc{gw}(f):=  \frac{1}{\rho_{\rm crit}} \frac{\rmd\rho_\textsc{gw}}{\rmd \ln f}=\frac{\pi^2f^2}{3H_0^2}\cP_{\rm t}(f)\,\cT^2(f)\,.
\ee
where $\rho_{\rm crit}:=3\Mpl^2H_0^2$ is the critical energy density, $\rho_\textsc{gw}:=[\Mpl^2/(8a^2)]\langle(\p_\tau h_{ij})^2+(\nabla h_{ij})^2\rangle$ is the energy density of GWs (spatial average of the kinetic energy of the transverse-traceless perturbation $h_{ij}$) and the shape of the transfer function $\cT$, which can be found in \cite{Kuroyanagi:2014nba}, depends on the expansion history of the universe. Substituting the values of cosmological parameters provided by \textsc{Planck} \cite{Akrami:2018odb} in the transfer function and making standard assumptions on the number of relativistic degrees of freedom in the early universe, the GW amplitude of the mode which enters the horizon during the radiation-dominated era can be written as 
\be\label{Omgw2}
\Om_\textsc{gw}(f)=\frac{\cA}{h^2}\,r\left(\frac{f}{f_0}\right)^{n_{\rm t}+\frac{\a_{\rm t}}{2}\ln\frac{f}{f_0}},\qquad \cA\approx 1.4\times 10^{-15}\,.
\ee
A blue-tilted spectrum can produce a stochastic GW background with increasing amplitude that can reach the sensitivity thresholds of GW interferometers at high frequencies.

%%%%%%%%%%%%%%%%%%%%%%%%%%%%%%%%%%%%%%%%%%%%%%%%%%%%%%%%%%%%%%%%%%%%%%%%%%%%%%%%%

\subsection{Results in quantum gravity}

For any given model, one can compare the theoretical prediction of the GW spectrum with the sensitivity curve of LIGO-Virgo-KAGRA \cite{TheLIGOScientific:2016wyq,Abbott:2017xzg,Akutsu:2018axf}, LISA \cite{Bartolo:2016ami,Caprini:2019pxz}, Einstein Telescope (ET) \cite{Maggiore:2019uih} and DECIGO \cite{Seto:2001qf,Kawamura:2011zz,Kawamura:2020pcg}, taking into account the cosmic microwave background (CMB) bound corresponding to $r<0.068$ and current and future sensitivity curves of pulsar timing experiments, NANOGrav \cite{Arzoumanian:2018saf} and SKA \cite{Janssen:2014dka}.

Obtaining a primordial blue-tilted tensor spectrum in quantum gravity is difficult. Despite the abundance of viable cosmological inflationary models in quantum gravity, a close scrutiny reveals that most of them predict a red tilt and those that have a blue tilt often lead to unobservable effects because $n_{\rm t}$ or $r$, or both, are too close to zero. 
\begin{itemize}
\item The large class of flux-compactification models in string cosmology is uniformly characterized by $n_{\rm t}<0$ \cite{CQC,Baumann:2014nda}. A case apart is the old ekpyrotic scenario, which predicts a strongly blue-tilted tensor index $n_{\rm t}=2$ \cite{KOST1}. While early versions of the model are ruled out because they have also a blue-tilted scalar spectrum, in a recent single-field version perturbations are generated before the ekpyrotic phase and the scalar spectrum is safely red-tilted \cite{KhSt1,KhSt2}. However, in all the realizations of the model the tensor-to-scalar ratio is exceptionally small and the resulting stochastic GW background is well below the detection threshold of any present or future interferometer \cite{Boyle:2003km}.
\item In Wheeler--DeWitt canonical quantum cosmology \cite{Kiefer:2011cc,Bini:2013fea,Brizuela:2016gnz,Kamenshchik:2017kfs,Kamenshchik:2015gua}, the semi-classical limit of the Wheeler--DeWitt equation for the wave-function of the Universe admits two solutions such that
the tensor spectrum is
\be\label{wdw} 
\cP_{\rm t}(k)\simeq\cP_{\rm t}^{(0)}(k) \left[1\pm c\,(\lp H)^2\left(\frac{k_0}{k}\right)^3\right],
\ee
where $\cP_{\rm t}^{(0)}(k)\propto H^2$ is the standard spectrum at horizon crossing ($k=aH$), $c>0$ is a known numerical constant, $\lp\approx 10^{-35}\,{\rm m}=5\times 10^{-58}\,{\rm Mpc}$ is the Planck length and $k_0$ is the pivot scale of the experiment: for the CMB, typically $k_0=0.05\,{\rm Mpc}^{-1}$ or $k_0=0.002\,{\rm Mpc}^{-1}$. The $-$ ($+$) sign corresponds to a blue (respectively, red) tilt. The strong suppression of the $(\lp H)^2\ll 1$ term is further increased at late times by the $(k_0/k)^3$ factor, since at the frequencies of LISA and DECIGO $k_0/k\sim 10^{-15}-10^{-13}$. Therefore, the quantum correction is unobservable.
\item In loop quantum cosmology, there are three main approaches to cosmological perturbations.
	\begin{itemize}
	\item In the dressed-metric approach, the tensor spectrum is red tilted \cite{Agullo:2015tca,Li:2019qzr}. 
	\item In the effective-constraints or anomaly-cancellation approach, one can consider quantum corrections coming from inverse-volume operators, from hol\-onomies or, more realistically, from both. If one considers only inverse-volume corrections, the inflationary spectra are compatible with observations but the tensor spectrum is red tilted \cite{BCT2,Zhu15}. The case with only holonomy corrections predicts a blue-tilted tensor spectrum but it is ruled out observationally \cite{BoBGS}. To the best of our knowledge, the case with both types of corrections has not been explored yet.
	\item In the hybrid-quantization approach, the value and sign of the spectral index $n_{\rm t}(k)$ depend on the background effective solution and, even more importantly, on the vacuum on which to perturb such background. The $k$-dependence of the tensor index can be found via a numerical analysis and it turns out that for some choices of vacuum $n_{\rm t}<0$, while for others the spectrum oscillates rapidly and a blue tilt can be generated at certain frequencies \cite{deBO,Gomar:2017yww}. However, what matters for the formation of a stochastic GW background is the average trend of the spectrum and, in all these cases, it decreases in $k$ when $k$ is sufficiently large. Therefore, the spectrum is red-tilted at small scales and it is unlikely that this model could generate a detectable stochastic background, for any vacuum choice. A detailed numerical study, which we will not pursue here, could give a more precise answer.
	\end{itemize}
\item Non-local quantum gravity offers a natural embedding of Starobinsky gravity into a fundamental theory. The early universe is described by a period of inflation driven by a non-local quadratic gravitational action \cite{Briscese:2013lna,Koshelev:2016xqb,Koshelev:2017tvv,Koshelev:2020foq,CaKu}. Decompose the metric $g_{\mu\nu}=g_{\mu\nu}^{(0)}+h_{\mu\nu}$, where $g_{\mu\nu}^{(0)}$ is a background (typically, FLRW) and $h_{\mu\nu}$ is an inhomogeneous tensor fluctuation (i.e., not including scalar modes) corresponding to the graviton particle at the quantum level. In the frame where the theory is defined, in terms of an action made of quadratic curvature invariants, the perturbed action on a quasi de Sitter background is
\be\label{S2}
S=\frac{\Mpl^2}{2}\int\rmd^4x\,\sqrt{|g|}\,h_{ij}\left(\B-\frac{R}{6}\right)\,\rme^{\tilde\H_2(\B)}h^{ij}\,,
\ee
where $g$ is the determinant of the background metric $g_{\mu\nu}^{(0)}$, $\B=\N_\mu\N^\mu$ and $R$ are, respectively, the Laplace--Beltrami operator and the Ricci scalar on the background, $\tilde\H_2(z):=\H_2(z-4z_*)$, $\H_2$ is a function such that the form factor $\exp(-\H_2)$ is entire (i.e, it does not contain any poles in the momentum) and $z_*=R/(6M_*^2)$, $M_*$ being the fundamental mass scale of the theory. The background accelerating solution is the same as in local Starobinsky inflation, so that on this quasi de Sitter background the slow-roll approximation holds. The approximate primordial tensor spectrum as a function of the comoving wave-number is 
\ba
\cP_{\rm t}(k) &\sim&\left[1-\frac{3}{2\ln(k_{\rm e}/k)}\right]\,\rme^{-\tilde\H_2\left[\frac{2H^2(k)}{M_*^2}\right]}\,,\label{Ptfull2}\\
H(k) &\sim&\left(\ln\frac{k_{\rm e}}{k}\right)^{\frac12}+\frac{1}{12}\left(\ln\frac{k_{\rm e}}{k}\right)^{-\frac12}\,,\label{Hk}
\ea
up to proportionality factors, where $k_{\rm e}$ is the wave-number of the last perturbation exiting the horizon at the end of inflation. Equation \Eq{Hk} tells us that the Hubble parameter is approximately proportional to a positive power of $\ln(k_{\rm e}/k)>0$ during inflation. Thus, both $H$ and $z_*\propto H^2$ decrease when $k$ increases. Since a characteristic of non-local quantum gravity is that $\tilde\H_2$ increases with $z_*$, then $\tilde\H_2$ decreases when $k$ increases and vanishes asymptotically. Therefore, at high frequencies (high $k$) the non-local term in the tensor spectrum \Eq{Ptfull2} tends to unity, $\lim_{k\to\infty}\exp[-\tilde\H_2(z_*)]=1$. Therefore, non-local Starobinsky gravity is out of reach of any present or future GW interferometer because at high frequencies it reduces to standard Starobinsky inflation, which has a red-tilted spectrum \cite{CaKu}.
\end{itemize}

Other models of, or related to, quantum gravity manage to generate a blue-tilted tensor spectrum and a detectable stochastic GW background: pre-big-bang cosmology, string-gas cosmology, new ekpyrotic scenario, Brandenberger--Ho non-commutative inflation and multi-fractional spacetimes. Except pre-big-bang cosmology, which will be mentioned only in the conclusions, we shall discuss them separately, while we will follow a model-independent approach for the luminosity distance. 

\subsubsection{String-gas cosmology}

In a compact space, the excitation modes of a thermal ensemble of strings are momentum modes and winding modes. The energy of winding modes decreases with the size of the available space and their number increases with the energy, so that, for an adiabatic process, winding modes dominate the thermal bath in a small space. The temperature of this bath cannot rise indefinitely and reaches a maximal temperature $T_{\rm H}$ called Hagedorn temperature. The universe starts with an almost constant scale factor and a temperature slightly lower than $T_{\rm H}$. Both scalar and tensor spectra are generated thermally. In particular, tensor modes are generated by anisotropic pressure terms in the energy-momentum tensor, but near the Hagedorn temperature the thermal bath is dominated by winding modes and the pressure decreases. Thus, there is a decrease of power at low $k$ and a slight blue tilt. Eventually, winding modes decay and three spatial directions open up while the others stay compact, leading to a radiation-dominated era.

In this scenario, inflation is replaced by a quasi-static era where thermal fluctuations generate an almost scale-invariant primordial scalar and tensor spectrum, the latter being 
\be\label{sgc2}
\cP_{\rm t}(k)\simeq\frac{1}{4(\Mpl l_{\rm st})^4}\hat T(k)\left[1-\hat T(k)\right]\,\ln^2\left[\frac{1-\hat T(k)}{l_{\rm st}^2k^2}\right],
\ee
where $l_{\rm st}$ is the string length scale, $\hat T(k):=T(k)/T_{\rm H}$ and the temperature $T(k)$ is evaluated at the time when the mode with comoving wave-number $k$ exits the horizon. The form of $T(k)$ is unknown except for its behaviour during the Hagedorn phase ($T\approx {\rm const}\lesssim T_{\rm H}$) and in the following radiation-domination era ($T\sim 1/a$). In the scalar sector, an increase of power is observed for small $k$ ($n_{\rm s}-1<0$), while the tensor index and its running read
\ba
n_{\rm t}&\simeq& (1-n_{\rm s})-\frac{4}{\ln[(1-\hat T)(l_{\rm st}k)^{-2}]}>0\,,\label{ntstga}\\
\a_{\rm t}&\simeq&-\a_{\rm s}\,.
\ea

A blue-tilted tensor spectrum is one of the characteristic predictions of string-gas cosmology that could be tested if a primordial gravitational signal was discovered. However, in order to achieve detection the small blue tilt \Eq{ntstga} $n_{\rm t}\approx 0.035$ \cite{Akrami:2018odb} should be accompanied with as high as possible a tensor-to-scalar ratio:
\be
r=\frac{25}{9}\left(1-\hat T\right)^2\ln^2\left[\frac{1-\hat T}{(l_{\rm st}k)^2}\right]\,.
\ee

\begin{figure}
\centering
\includegraphics[width=10cm]{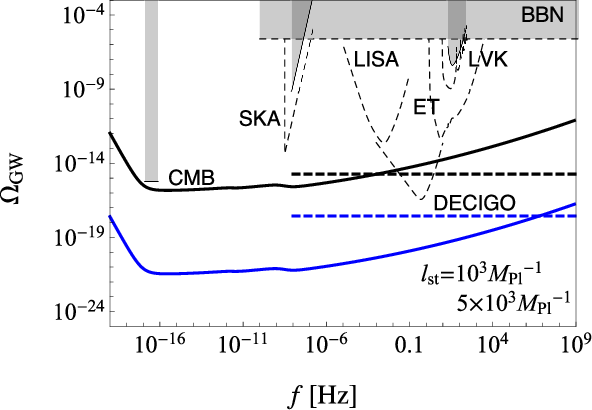}
\caption{\label{fig1} Stochastic GW background of string-gas cosmology compared with the sensitivity curves of LIGO-Virgo-KAGRA (LVK), SKA, LISA, ET and DECIGO. The black and blue solid curves correspond to $l_{\rm st}=10^3 \Mpl^{-1}$ and $5\times 10^3 \Mpl^{-1}$, respectively. We used the central values of the \textsc{Planck} constraints $n_{\rm s}=0.9658$ and $\a_{\rm s}=-0.0066$. Note that the approximation \Eq{ptk} breaks down for high-frequency GWs, since the corresponding curves go beyond the upper bound on the GW amplitude (dashed lines) obtained from the full expression \Eq{sgc2}. Credit: \cite{CaKu}.}
\end{figure}

The stochastic GW background predicted by string-gas cosmology is shown in Fig.\ \ref{fig1}. In the most optimistic case of a tensor-to-scalar ratio saturating the CMB bound, the model reaches DECIGO sensitivity.

\subsubsection{New ekpyrotic scenario}

In the ekpyrotic universe, two flat 3-branes constitute the boundary of a five-dimensional spacetime and interact with an attractive potential $V(\vp)$ along a compact fifth dimension parametrized by the radion $\vp$. As the branes get closer, the gravitational energy in the bulk is converted into brane kinetic energy and the branes collide and oscillate back and forth their center of mass along the extra direction. During the collision, part of the brane kinetic energy is converted into matter and radiation. An observer on one of the branes experiences the brane collision as a big bang after a period of contraction called ekpyrosis. During the slow contraction phase, a pattern of inhomogeneities is developed. In the new scenario of \cite{Brandenberger:2020tcr,Brandenberger:2020eyf}, the tensor index is small and positive,
\be
n_{\rm t}=1-n_{\rm s}>0\,,\qquad \a_{\rm t}=-\a_{\rm s}\,.\label{ntekpy}
\ee
The tensor-to-scalar ratio at the pivot scale is
\be\label{rekpy}
r(k_0)\simeq \frac{2^{n_{\rm s}}25}{\Gamma^2\left(1-\frac{n_{\rm s}}{2}\right)}(k_0\tau_{\rm B})^{2(1-n_{\rm s})}\,(1-n_{\rm s})^2\,,
\ee
where $\Gamma$ is Euler's function and $\tau_{\rm B}$ is the conformal time at which the string density is reached and the cosmological bounce takes place. Taking the grand-unification scale $\tau_{\rm B}=(10^{16}\,{\rm GeV})^{-1}\approx 5\times 10^{-55}\,{\rm Mpc}$ and the observed value of the scalar index $n_{\rm s}\approx0.9658$, one has $r(k_0)\sim (10^{-4}- 10^{-2})\,(1-n_{\rm s})^2$.

\begin{figure}
\centering
\includegraphics[width=10cm]{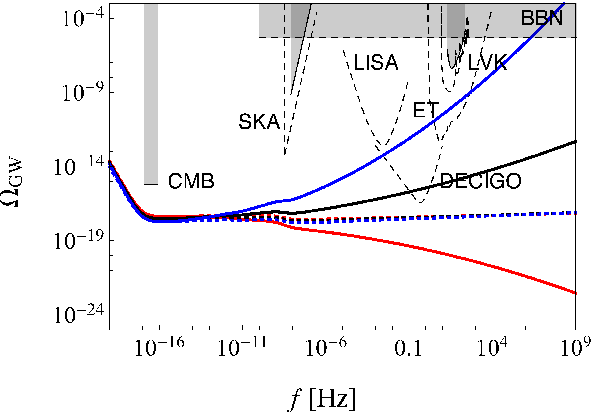}
\caption{\label{fig2} Stochastic GW background of the new ekpyrotic scenario compared with the sensitivity curves of LIGO-Virgo-KAGRA, SKA, LISA, ET and DECIGO. Denoting as $n_{\rm s}^{\rm obs}\pm\de n_{\rm s}$ and $\a_{\rm s}^{\rm obs}\pm\de \a_{\rm s}$ the \textsc{Planck} values, we plot the worst case minimizing the tensor blue tilt and maximizing the negative tensor running at the $2\s$-level ($n_{\rm t}=1-(n_{\rm s}^{\rm obs}+2\de n_{\rm s})\approx 0.026$, $\a_{\rm t}=-(\a_{\rm s}^{\rm obs}+2\de \a_{\rm s})\approx -0.007$, red solid curve), the intermediate case taking the central values of the parameters ($n_{\rm t}=1-n_{\rm s}^{\rm obs}\approx 0.034$, $\a_{\rm t}=-\a_{\rm s}^{\rm obs}\approx 0.007$, black solid curve) and the best case maximizing the tensor blue tilt and the positive tensor running at the $2\s$-level ($n_{\rm t}=1-(n_{\rm s}^{\rm obs}-2\de n_{\rm s})\approx 0.042$, $\a_{\rm t}=-(\a_{\rm s}^{\rm obs}-2\de \a_{\rm s})\approx 0.021$, blue solid curve). We take $\tau_{\rm B} = 5.8\times 10^{-24}\,{\rm Mpc}$, which corresponds to the conformal time at the grand-unification scale $10^{16}$\,GeV. The dotted curves correspond to the above cases with no running. Credit: \cite{CaKu}.}
\end{figure}

Figure \ref{fig2} shows the stochastic GW background of this model. The signal reaches the DECIGO curve only in the most optimistic case of a strong blue tilt and a strong positive tensor running.

\subsubsection{Brandenberger--Ho non-commutative inflation}

The graviton action of this model is decorated with *-products. This structure alters the scalar and tensor primordial spectra in a way that depends on whether the perturbation modes have a wave-length smaller (ultraviolet limit, UV) or larger (infrared limit, IR) than the fundamental length scale $\ell_{\rm nc}$ appearing in the commutation algebra. In the particular case of natural inflation, the IR limit of this model has a blue-tilted tensor spectrum and is compatible with \textsc{Planck} data \cite{Calcagni:2013lya}. Here the mechanism is directly related to the time-momentum uncertainty, which induces a $k$-dependence in the effective mass term of the GW propagation equation. This dependence is inherited by the tensor spectrum and, for a certain choice of parameters, it leads to an enhancement at small scales. In the IR limit,
\be\label{Anoncom}
\cP_{\rm t} = \cP^{(0)}_{\rm t}\,\Sigma^2 (\ell_{\rm nc} H)\,,
\ee
where $\cP_{\rm t}^{(0)}=\cP_{\rm t}(\Sigma\!\!=\!\!1)$ is the amplitude in the commutative limit (Einstein gravity) and $\Sigma$ is a function encoding the non-commutative effects. $\Sigma$ multiplies both the tensor and scalar amplitudes, so that their ratio $r$ is unchanged. When the FLRW 2-sphere is factored out of the total measure, the tensor spectral index is positive,
\be
n_{\rm t}\simeq \frac{r}{4}\simeq 4\e,\qquad \a_{\rm t} \simeq 8\e(\e-\eta)\,,\label{nt0}
\ee
where $\e:=-\dot H/H^2$ and $\eta:= -\ddot{\phi}/(H \dot\phi)$ are the first two slow-roll parameters. The scalar spectrum is red-tilted for some choices of inflaton potential, such as natural inflation \cite{Calcagni:2013lya}. The dependence of the observables \Eq{nt0} from the number of e-foldings $\cN$ and the parameter $A=(\phi_*/\Mpl)^2/3$, where $\phi_*$ is a characteristic scale of the scalar-field potential, can be found in \cite{Calcagni:2013lya,CaKu}. In particular, CMB data constrain $5.8<A<11$ at the 95\,\% CL for $50<\cN<60$.

As expected by the fact that non-commutativ\-i\-ty changes the sign and value of the coefficients in the slow-roll expressions of the observables but not their order of magnitude, the spectrum is enhanced up to the DECIGO sensitivity curve, but barely so, and it does not reach ET \cite{CaKu}.

\subsubsection{Multi-fractional spacetimes}

Multi-fractional spacetimes are spacetimes where the clocks and rulers used by the observer register different scaling laws (for instance, linear size versus volume) in copies of the same object with  different sizes (for instance, a human-size hyperball compared with a microscopic one) \cite{revmu}. This ever-changing geometry is typical of spacetimes arising in quantum gravity \cite{tH93,Car17}. Multi-fractional theories implement this dimensional flow via a modification of the integration measure in the field action and of the kinetic operators acting on the fields, including gravity. In the particular case of the so-called theory with $q$-derivatives, the tensor spectrum reads
\be\label{Pt}
\cP_{\rm t}(k)=r(k_0)\,\cP_{\rm s}(k_0)\,\exp\left\{n_{\rm t}(k_0)\,\ln\frac{p(k)}{p(k_0)}+\frac{\a_{\rm t}(k_0)}{2}\left[\ln\frac{p(k)}{p(k_0)}\right]^2\right\}\,,
\ee
where $n_{\rm t}<0$, $\a_{\rm t}<0$ and $r=-8n_{\rm t}$ as in standard inflationary models and
\be
p(k) \simeq k\left[1+\frac{1}{|\a|}\left(\frac{k}{k_*}\right)^{1-\a}\right]^{-1},\label{Pt2}
\ee
where $k_*$ is a fundamental comoving scale and the parameter $\a$ (not to be confused with the running of scalar or tensor indices) is related to the Hausdorff dimension of space (the way volumes scale with the linear size; see below) by $\dh^{\rm space}=3\a$ in three topological dimensions. If $0<\a<1$ (respectively, $\a>1$), the tensor spectrum is red-tilted but less (respectively, more) than in Einstein gravity. If $\a<0$, for $k\ll k_*$ (large scales, small frequencies) one recovers GR, while an exotic regime is reached for $k\gg k_*$ (small scales, large frequencies) where $p(k)\simeq k^\a$ and $P_{\rm t}(k)\sim k^{\a n_{\rm t}+\frac12\a_{\rm t}\a^2\ln(k/k_0)}$. The observed spectral index $\a n_{\rm t}$ is positive definite and the spectrum increases at large $k$, while the observed tensor running at large frequencies stays negative, $\a^2\a_{\rm t}<0$. The geometric interpretation is that in the IR the Hausdorff dimension of spacetime is smaller than 4 and there is an increase in the number of modes given the same density. 

\begin{figure}
\centering
\includegraphics[width=10cm]{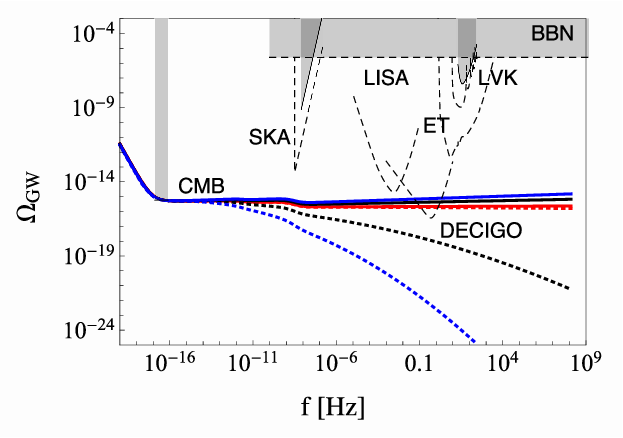}
\caption{\label{fig3} Stochastic GW background of multi-fractional inflation with no running ($\a_{\rm t}=0$) and $\a=-1/2,-3,-5$ (respectively, red, black and blue solid curve), compared with the sensitivity curves of  LIGO-Virgo-KAGRA, SKA, LISA, ET and DECIGO. Here $r=0.06$ and $n_{\rm t}\approx -0.0075$ is given by the consistency relation $r=-8n_{\rm t}$. The $\a_{\rm t}=0$ plots are unaffected by different choices of $k_*$, which we took in the range $k_*=10^{-20}-10^{10}\,{\rm Mpc}^{-1}$. The dotted curves correspond to the above cases with non-zero running $\a_{\rm t}=-0.0001$ (typical order of magnitude of inflationary models) and $k_*=10^{-3}\,{\rm Mpc}^{-1}$. Credit: \cite{CaKu}.}
\end{figure}

As one can see in Fig.\ \ref{fig3}, for the typical signs and values of $r$, $n_{\rm t}$ and $\a_{\rm t}$ of standard inflationary scalar-field models, the theory can reach the DECIGO sensitivity curve if the tensor-to-scalar ratio $r$ is close to the CMB bound (in the figure, $r=0.06$). The running decreases the spectral amplitude at the typical frequencies of interferometers. Results with zero running are insensitive of the choice of scale $k_*$ (position of the bending point of the spectrum) because $\a n_{\rm t}$ is very small, while there is a visible effect in the presence of running due to the amplification by $\a^2$.

%%%%%%%%%%%%%%%%%%%%%%%%%%%%%%%%%%%%%%%%%%%%%%%%%%%%%%%%%%%%%%%%%%%%%%%%%%%%%%%%%
%%%%%%%%%%%%%%%%%%%%%%%%%%%%%%%%%%%%%%%%%%%%%%%%%%%%%%%%%%%%%%%%%%%%%%%%%%%%%%%%%

\section{Modified dispersion relation and propagation speed}

%%%%%%%%%%%%%%%%%%%%%%%%%%%%%%%%%%%%%%%%%%%%%%%%%%%%%%%%%%%%%%%%%%%%%%%%%%%%%%%%%

\subsection{Basics}

A feature common to all theories of quantum gravity is dimensional flow, the change of spacetime dimension with the probed scale. Quantization of spacetime geometry or its emergence from fundamental physics introduces, directly or indirectly, two types of change relevant for the propagation of GWs: an anomalous spacetime measure $\rmd\vr(x)$ (how volumes scales) and a kinetic operator $\cK(\p)$ (modified dispersion relations), so that the perturbed action for a small perturbation $h_{\mu\nu}$ over a background $g^{(0)}_{\mu\nu}$ is
\be\label{hboxh2} \hspace{-.5cm}
S=\frac{1}{2\ell_*^{2\Gamma}}\!\!\int\!\rmd\vr\sqrt{|g^{(0)}|}
\left[h_{\mu\nu}\cK h^{\mu\nu}\!+O(h_{\mu\nu}^2)\right]\!, 
\ee
up to a source term, where $\ell_*$ is a (or the) characteristic length scale of the geometry and $\Gamma$ is a constant. Splitting the perturbation into the usual polarization decomposition $h_{\mu\nu}=h_+e^+_{\mu\nu}+h_\times e^\times_{\mu\nu}$, the modes $h_{+,\times}/\ell_*^{\Gamma}$ are dimensionally and dynamically equivalent to a scalar field.
	
The measure defines a geometric observable, the Hausdorff dimension $\dh(\ell):={\rmd\ln\vr(\ell)}/{\rmd\ln\ell}$,
describing how volumes scale with their linear size $\ell$. In a classical spacetime, $\dh=D$. Also, spacetime is dual to a well-defined momentum space characterized by a measure $\tilde\vr(k)$ with Hausdorff dimension $\dh^k$, in general different from $\dh$. The kinetic term
is related to $\dh^k$ and to another observable, the spectral dimension $\ds(\ell):=-{\rmd\ln P(\ell)}/{\rmd\ln\ell}$, where $P(\ell)\propto\int\tilde\vr(k)\,\exp[-\ell^2\tilde\cK(-k^2)]$. In any plateau of dimensional flow, where all dimensions are approximately constant, $\tilde\vr(k)\sim\rmd k\,k^{\dh^{k}-1}$ and 
\be\label{dr1}
\tilde\cK(-k^2)=-\ell_*^{2-2\b}k^2+k^{2\b}\,,
\ee
where $\b:=[\cK]/2$ is a constant given by half the energy scaling of $\cK$. Here momentum-space coordinates have the usual dimensional units $[k^\mu]=1$. Then, one finds that $P\propto(\ell_*^{\b-1}\ell)^{-\dh^{k}/\b}$, implying $\ds=2\dh^k/[\cK]$ and
\be
\b=\frac{\dh^k}{\ds}\,.
\ee
In such a plateau region, since $[S]=0$, from \Eq{hboxh2} we have  
\be\label{vp}
\Gamma\simeq\frac{\dh}{2}-\frac{\dh^k}{\ds}\,,
\ee
and $\Gamma\approx{\rm const}$. We assume that $\ds\neq 0$ at all scales. The case of non-local quantum gravity, where $\ds=0$ at short scales, must be treated separately. In the GR limit in $D$ topological dimensions, $\dh=\dh^k=\ds=D$ and $\Gamma=D/2-1$, the usual scaling of a scalar field. In $D=4$, $\G=1$. The value of the dimensions $\dh$, $\dh^k$ and $\ds$ and of the parameter $\Gamma$ for different quantum gravities can be found in \cite{Calcagni:2019kzo}; in the UV, $-3\leq \G_{\rm UV}\leq 2$.

The modified dispersion relation \Eq{dr1} arises in higher-order non-unitary theories ($\b=2,3,\dots$) \cite{Ste77,ALS,AAM}, the propagation of low-energy particles in non-critical string theory ($\b=3/2$ at mesoscopic scales and $\b=2$ in the deep UV) \cite{ACEMN}, asymptotic safety ($\b=2$) \cite{LaR5}, Ho\v{r}ava--Lifshitz gravity ($\b=3$ only in the spatial directions) \cite{Hor3}, multi-fractional spacetimes with fractional and $q$-derivatives ($\b>0$; Lorentz invariance is broken here) \cite{revmu}, causal sets ($\b=2$) \cite{BBL} and as an effective dispersion relation in loop quantum gravity \cite{GaPu,AMTU,ACAP,Ron16}. Last, $\ell_*$ is a length scale around which quantum-gravity effects become relevant. If there is only one fundamental scale $\ell_*$, then it is very small (of order of, or not too far from, the Planck scale $\lp$) and the UV regime corresponds to very short wave-lengths. If, however, \Eq{dr1} is an effective dispersion relation, $\ell_*$ is a mesoscopic scale that could be relatively far from the actual UV and closer to the IR regime.

%%%%%%%%%%%%%%%%%%%%%%%%%%%%%%%%%%%%%%%%%%%%%%%%%%%%%%%%%%%%%%%%%%%%%%%%%%%%%%%%%

\subsection{Results in quantum gravity}

The dispersion relation \Eq{dr1} of the spin-2 graviton field has been used to impose constraints on quantum-gravity theories exhibiting dimensional flow using the LIGO-Virgo merging events \cite{Yunes:2016jcc,EMNan,ArCa2}. Given the dispersion relation \Eq{dr1}, the velocity of propagation of a wave front is given by the group velocity
\be\label{groupv}
c_\textsc{gw}:=\frac{\rmd\om}{\rmd |\mathbf{k}|}\,.
\ee
The signal of GW150914 is peaked at a frequency $f=\om/(2\pi)=100\,{\rm Hz}$, corresponding to $\om\approx 630\,{\rm Hz}\approx 4.1\times 10^{-13}\,{\rm eV}$. At that frequency, the difference $\De c:= c_\textsc{gw}-1$ between the propagation speed of the signal and the speed of light is bounded from above as \cite{TheLIGOScientific:2016src}
\be\label{devbo}
|\De c| < 4.2\times 10^{-20}\,.
\ee
For the usual Lorentz-invariant massive dispersion relation $\om^2=|\mathbf{k}|^2+m^2$, one gets $c_\textsc{gw}\simeq 1-m^2/(2\om^2)$ if the mass is small, so that $\De c\simeq -m^2/(2\om^2)$ and $m<1.2\times 10^{-22}\,{\rm eV}$. Consider now the more general dispersion relation $\om^2 = |\mathbf{k}|^2 +O(1)\ell_*^n|\mathbf{k}|^{n+2}$. The case $n=1$ stems from generic quantum-gravity arguments \cite{ACEMN}, while the case $n=2$ can be obtained either as the low-momentum expansion of Padmanabhan's non-local model or from an argument matching the logarithmic leading-order LQG correction to the entropy-area law for black holes \cite{ACAP}. Since the GW frequency is much lower than the Planck frequency, one gets very weak bounds on $\ell_*$ when $n=1,2$ \cite{EMNan,ArCa2}. However, setting instead $\ell_*^{-1}>10\,{\rm TeV}$ (quantum-gravity scale larger than the LHC scale), one finds the bound $0<n<0.76$ \cite{ArCa2}. We can connect these results with the kinetic operator \Eq{dr1} if we allow for a breaking of Lorentz invariance. Expanding the dispersion relation $\tilde\cK=0$ for small $\om/|\mathbf{k}|$, one finds
\be
\om^2=|\mathbf{k}|^2-\ell_*^{2\b-2}|\mathbf{k}|^{2\b}\left(1-\frac{\om^2}{|\mathbf{k}|^2}\right)^\b\simeq |\mathbf{k}|^2-\ell_*^{2\b-2}|\mathbf{k}|^{2\b},
\ee
where $n=2\b-2$. As said, too weak bounds on $\ell_*$ are obtained when $n=1,2$ ($\b=3/2, 2$), while when $\ell_*^{-1}>10\,{\rm TeV}$ we have $\b=\dh/2-\G<1.38$ when $\dh\approx 4$. This happens at intermediate scales where the corrections to GR are small but non-negligible. In this mesoscopic regime, the above bound implies 
\be\label{bouw}
\G_{\rm meso}>0.62\,,
\ee
weaker than the constraint \Eq{bou1} discussed below. At any rate, one should apply this bound with care to Lorentz-invariant theories, as it entailed an assumption ($\om/|\mathbf{k}|\ll 1$) that may strongly depend on the model and on the symmetry-breaking mechanism.

Multi-fractional spacetimes can produce \emph{ad hoc} models \cite{revmu,Yunes:2016jcc} saturating the bound \Eq{bouw} which, in turn, would constrain the two-dimensional parameter space $(\ell_*,\a)$ of the simplest version of the theory to a region that should be checked against other types of cosmological observations and mechanisms (inflation, dark energy, and so on). While this procedure can select specific viable models, it is purely phenomenological.

The modified dispersion relation of the graviton in non-local quantum gravity requires a separate but rather quick analysis. The linearized perturbation equation is \cite{Modesto:2011kw}
\be\label{liper}
\B\tilde h=0\,,\qquad \tilde h:= \rme^{\H_2(\B)} h\,,
\ee
where $\H_2$ is a non-local form factor such that $\exp\H_2$ is an entire function of the background Laplace--Beltrami operator $\B$. The dispersion relation is $k^2\exp\H_2(-k^2)$ $=0$, which leads to the usual pole $k^2=-\om^2+|\mathbf{k}|^2=0$ in the propagator. Since the perturbation equation is the standard one for $\tilde h$, the propagation speed of GWs equals the speed of light and the theory avoids the bound \Eq{devbo}.

Early-universe scenarios such as Brandenberger--Ho non-commutative inflation %, string-gas cosmology
 and the new ekpyrotic scenario have no say about the propagation of late-time GWs. Therefore, they are not constrained by interferometric observations on the propagation speed or the luminosity distance of individual sources.

We conclude that, in general, modified dispersion relations in quantum gravity are not an efficient way to constrain the theory because the correction is too small and the graviton speed is very close, or equal to, the speed of light, thus evading the bound \Eq{devbo}.

%%%%%%%%%%%%%%%%%%%%%%%%%%%%%%%%%%%%%%%%%%%%%%%%%%%%%%%%%%%%%%%%%%%%%%%%%%%%%%%%%
%%%%%%%%%%%%%%%%%%%%%%%%%%%%%%%%%%%%%%%%%%%%%%%%%%%%%%%%%%%%%%%%%%%%%%%%%%%%%%%%%

\section{Luminosity distance}

%%%%%%%%%%%%%%%%%%%%%%%%%%%%%%%%%%%%%%%%%%%%%%%%%%%%%%%%%%%%%%%%%%%%%%%%%%%%%%%%%

\subsection{Basics}

The luminosity distance $d_L^\textsc{em}$ of a source of electromagnetic radiation is defined by the relation between the flux ${\rm F}$ of light reaching an observer and the power L per unit area emitted by the source:
\be\label{ludi}
{\rm F}=:\frac{{\rm L}}{4\pi (d_L^\textsc{em})^2}\,.
\ee
In standard GR, the luminosity distance as a function of the redshift $1+z=a_0/a$ (with $a_0=1$) is
\be\label{dLGR}
d_L^\textsc{em}(z)=(1+z)\int^{t_0}_{t(z)}\frac{\rmd t}{a}=(1+z)\int_0^z\frac{\rmd z}{H_\textsc{gr}}\,,
\ee
where $a(z)=(1+z)^{-1}$. The Hubble parameter $H_\textsc{gr}(z)$ is determined by the first Friedmann equation $H_\textsc{gr}^2=(\k^2/3)\rho$ and includes all energy contributions, including dark-energy. At small $z$, $d_L^\textsc{em}\simeq z/H_0$, where $H_0$ is the Hubble parameter today.

Sources of GWs admit another definition of luminosity distance. Let $h_{\mu\nu}$ be a metric perturbation around the Minkowski background $\eta_{\mu\nu}={\rm diag}(-,+,\cdots,+)$ and call $h$ one of the graviton polarization modes. The scalar $h$ is the amplitude of a gravitational wave emitted by a source such as a black-hole or a neutron-star binary system. In $D$ topological dimensions, a direct calculation \cite{CDL} or a scaling argument \cite{Calcagni:2019ngc} yields the asymptotic dependence of $h$ on the distance $r=|{\bf x}-{\bf x}'|$ in the local wave zone of the source, i.e., a region of space much larger than the wave-length of the metric perturbation but smaller than cosmological scales:
\[
h\simeq \frac{\k\cF_h(t-r)}{r^{\frac{D-2}{2}}}\stackrel{D=4}{\propto}\frac{1}{r}\,,
\]
where $\cF_h$ is a function of retarded time. If the observer is at a cosmological distance, the cosmic expansion must be taken into account. Since $r$ is nothing but the comoving distance from the source, after rescaling with the scale factor $a$ one gets (see \cite[section 4.1.4]{Mag07} for the derivation in four dimensions)
\be\label{hcos}
h\sim \frac{1}{(d_L^\textsc{em})^{\frac{D-2}{2}}}\stackrel{D=4}{=}\frac{1}{d_L^\textsc{em}}\,.
\ee
For sources of GWs and light, called standard sirens, both sides of this equation can be measured: the left-hand side is the strain measured in an interferometer, while the right-hand side is determined by observations in the optical spectrum.

In theories beyond GR, the relation between $h$ and $d_L^\textsc{em}$ can be different. Defining the GW luminosity distance as $h=:1/d_L^\textsc{gw}$ (up to a retarded time-dependent function), in these theories the ratio $d_L^\textsc{gw}(z)/d_L^\textsc{em}(z)$ deviates from 1. This is a cosmological observable that can be determined from standard-sirens data and is parametrized in several forms \cite{Belgacem:2019pkk}.

%%%%%%%%%%%%%%%%%%%%%%%%%%%%%%%%%%%%%%%%%%%%%%%%%%%%%%%%%%%%%%%%%%%%%%%%%%%%%%%%%

\subsection{Results in quantum gravity}

To extract the luminosity distance in quantum gravity in a model-independent way, we appeal again to dimensional flow. In a multi-scale spacetime such as those arising in quantum gravity, the measurement of a generic distance $r$ in the absence of curvature in a non-relativistic regime (Newtonian approximation) deviates from the one in ordinary space by a power-law correction, so that \cite{NgDa,Ame94,ACCR,revmu}
\be\label{L}
r\to\tilde r = r\left[1+\e\left(\frac{r}{l_*}\right)^{\a-1}\right],
\ee
where $\a$ is the same as in \Eq{Pt2} and the parameter $\e$ accounts for two types of corrections (deterministic if $\e=\pm 1$, stochastic if $\e$ is a random variable averaging to zero \cite{ACCR}). According to the flow-equation theorem \cite{revmu}, when measuring $r$ with physical rods one can identify $\a$ with the UV Hausdorff dimension of spacetime divided by $D$ \cite{ACCR}. Generic quantum-gravity arguments select the values $\a=1/3,\,1/2$ as especially appealing \cite{NgDa,Ame94,ACCR}, although they are not really preferred in most of the concrete theories listed here.

It turns out that the luminosity distance follows a similar multi-scale power law. It is not difficult to show that \Eq{vp} is the scaling of the GW amplitude $h$ (subscripts $+,\times$ omitted) with respect to the radial distance $r$ in the local wave zone, $h(t,r)\sim f_h(t,r)\,(\ell_*/r)^{\G}$. On cosmological distances, it is sufficient to replace $r\to ar$. Assuming that quantum-gravity corrections to $d_L^\textsc{em}$ are negligible at  large scales and absorbing redshift factors and all the details of the source (chirp mass, spin, and so on) into the dimensionless function $f_h(z)$, one has
\[
h(z)\sim f_h(z)\,\left[\frac{\ell_*}{d_L^\textsc{em}(z)}\right]^{\G}. %f_h(z)\,\left[{\ell_*}/{d_L^\textsc{em}(z)}\right]^{\G}.
\] 
The final step is to generalize this relation, valid only for a plateau in dimensional flow, to all scales. An exact calculation is extremely difficult except in special cases, but a model-independent approximate generalization is possible because the system is multi-scale (it has at least an IR and a UV limit, respectively $\G\to 1$ and $\G\to\G_{\rm UV}$). In fact, multi-scale systems such as those in multi-fractal geometry, chaos theory, transport theory, financial mathematics, biology and machine learning are characterized by at least two critical exponents $\G_1$ and $\G_2$ combined together as a sum of two terms $r^{\G_1}+ A \, r^{\G_2}+\dots$, where $A$ and each subsequent coefficients contain a scale (hence the term multi-scale). In quantum gravity, lengths have exactly this behavior, which has been proven to be universal \cite{NgDa,Ame94,revmu,ACCR} in the flat-space limit. In particular, it must hold also for the luminosity distance because one should recover such a feature in the sub-cosmological limit $d_L^\textsc{em}\to r$. Thus \cite{Calcagni:2019ngc},
\be\label{dla} 
\frac{d_L^\textsc{gw}}{d_L^\textsc{em}}=1\pm|\g-1|\left(\frac{d_L^\textsc{em}}{\ell_*}\right)^{\g-1}, 
\ee 
with $\gamma\neq 0$. In the presence of only one fundamental length scale $\ell_*=O(\lp)$, \Eq{dla} is exact and $\g=\Gamma_{\rm UV}$ takes values different from 1, in the interval $-3\leq\Gamma_{\rm UV}\leq 2$ for the theories considered in this chapter. Conversely, if $\ell_*$ is a mesoscopic scale much larger than the Planck scale, then \Eq{dla} is valid only near the IR, close to the end of the flow, where $\g=\Gamma_{\rm meso}\approx 1$. This equation resembles the GW luminosity-distance relation expected in some models with large extra dimensions, where gravitons leak into a higher-dimensional space \cite{DeMe,PFHS,Andriot:2017oaz,Abb18}. 

Observations can place bounds on the two parameters $\ell_*$ and $\g$ in a model-independent way, by constraining the ratio \Eq{dla} as a function of the redshift of the source. An analysis based on two standard sirens (the binary neutron-star merger GW170817 observed by LIGO-Virgo and the Fermi telescope \cite{Ab17b} and a simulated $z=2$ super-massive black hole merging event that could be observed by LISA) shows that no constraint can be placed on the deep UV limit of any quantum gravity unless \Eq{dla} were valid at all scales and $0<\g-1=\G_{\rm UV}-1=O(1)$, in which case $\k$-Minkowski spacetime and Padmanabhan's non-local effective model would be ruled out. The other alternative is that \Eq{dla} was valid in a near-IR regime and $\g=\G_{\rm meso}$ was very close to 1 from above, in which case one finds the bound \cite{Calcagni:2019ngc}
\be\label{bou1}
0\,<\,\Gamma_{\rm meso}-1\,<\,0.02\qquad (\ell_*=\lp)\,.
\ee
Examining \Eq{vp}, one concludes that this case is realized only for geometries with a spectral dimension reaching $\ds\to
4$ from above. The only theories in our list that do so are those where $\G_{\rm UV}>\G_{\rm meso}>1$ ($\k$-Minkowski spacetime with
ordinary measure and bicross-product or relative-locality Laplacians and Padmanabhan's model) or $\G_{\rm meso}>1>\G_{\rm UV}$ (LQG/SF/GFT).
One can exclude observability of the models with $\G_{\rm UV}>\G_{\rm meso}>1$, since they predict $\G_{\rm meso}-1\sim (\lp/d_L^\textsc{em})^2< 10^{-116}$. Thus, only LQG/SF/GFT could generate a signal detectable with standard sirens, unless some yet unknown theoretical constraints limited the size of the effect. Here $\ds$ runs from small values in the UV, but before reaching the limit $\ds^{\rm IR}=4$ it overshoots the asymptote and decreases again \cite{COT3}: hence $\G_{\rm meso}>1>\G_{\rm UV}$. 

A complementary solar-system constraint on the spin-2 sector can arise from modifications of Newton's potential and can be much stronger than \Eq{bou1}, but it heavily relies on model-dependent assumptions on the scalar perturbation sector which are under poor control \cite{Calcagni:2019kzo}. On the other hand, the bounds obtained from $d_L$ are stronger than the ones found from the modified dispersion relation \Eq{dr1}. This is one reason behind the recent surge of interest in the luminosity distance to probe theories beyond GR.

Let us now discuss the case of non-local quantum gravity \cite{BrCM}. Using the same full calculation or the scaling argument as in GR with $h$ replaced by $\tilde h$, for entire form factors we have
\be
\tilde h= \frac{1}{d_L^\textsc{gw}}\qquad\Longrightarrow\qquad h= \rme^{-\H_2} \frac{1}{d_L^\textsc{gw}}\,,
\ee
where $\tilde h$ is defined in \Eq{liper}. We can estimate the non-local correction in the right-hand side for the minimal form factor $\H_2(\B)=-\ell_*^2\B=\ell_*^2(\p_t^2+3H\p_t)$ in the homogeneous approximation and, crudely, an approximately constant Hubble parameter $H\simeq H_0$, so that at large redshift $h\simeq H_0\rme^{-10(\ell_*H_0)^2}\rme^{2H_0(t-t_0)}$, while at small redshift $h\simeq H_0\rme^{-3(\ell_*H_0)^2}\rme^{2H_0(t-t_0)}$. Overall,
\be
h\simeq \frac{\rme^{-c (\ell_*H_0)^2}}{d_L^\textsc{gw}}\,,\qquad c=O(1)-O(10)\,.
\ee
Assuming, to maximize the effect, that light is not affected by non-locality, we have
\be\label{Calcdd}
\frac{d_L^\textsc{gw}}{d_L^\textsc{em}}\simeq 1+c (\ell_*H_0)^2\,,
\ee
and, for $\ell_*=\lp$, the right-hand side is of order of $1+10^{-120}$, an effect completely unobservable compared with the estimated error $\Delta d_L/d_L\sim 0.001-0.1$ of present and future interferometers \cite{Dalal:2006qt,Nissanke:2009kt,Camera:2013xfa,Tamanini:2016zlh}. For a power-law expansion $a=(t/t_0)^p$, $d_L\propto (t_0/t)^{2p}(t_0-t)$ and one can show that, again, the correction in the ratio \Eq{Calcdd} is of the order of $(\ell_*/t_0)^2\sim 10^{-120}$. Increasing $\ell_*$ to particle-physics scales does not magnify this correction enough, since it is governed by the cosmological scale $H_0^{-1}\sim t_0\sim 10^{17}\,{\rm s}$. Therefore, no non-local effect is observable in the luminosity distance for this theory.

Regarding multi-fractional spacetimes, it is possible to construct models with large deviations from the standard luminosity distance \cite{Calcagni:2019kzo} but, just as in the case of the modified dispersion relation, one enters the realm of \emph{ad hoc} phenomenology.

%%%%%%%%%%%%%%%%%%%%%%%%%%%%%%%%%%%%%%%%%%%%%%%%%%%%%%%%%%%%%%%%%%%%%%%%%%%%%%%%%
%%%%%%%%%%%%%%%%%%%%%%%%%%%%%%%%%%%%%%%%%%%%%%%%%%%%%%%%%%%%%%%%%%%%%%%%%%%%%%%%%

\section{Strain noise}

The above constraints can be complemented by a bound on the Hausdorff dimension of spacetime in the UV ($\dh^{\rm UV}$) coming from the strain noise of interferometers \cite{Ame98,Ame13,ACCR}. The correction in \Eq{L} can also be regarded as a threshold on the minimal uncertainty in physical measurements (spacetime fuzziness) of distances. Quantum gravity may manifest itself as an intrinsic noise with variance
\be\label{sqg}
\s_{\rm QG}^2=\ell_*^2\left(\frac{L}{\ell_*}\right)^{2\a},
\ee
where $\a$ is the same parameter introduced in \Eq{Pt2} and \Eq{L}. If $L$ is the typical length of a GW interferometer (e.g., the linear size of its arms), we can compare this quantum-gravity noise with the instrumental or strain noise $\s^2_{\rm exp}=\int\rmd f\,\cS^2(f)$ of a GW interferometer, where $\cS$ is the spectral noise. For a signal dominated by the characteristic frequency $1/L$ (in $c=1$ units), a rough estimate is $\s^2_{\rm exp}\simeq f\,\cS^2(f)|_{f=1/L}$. The strain noise is dimensionless and \Eq{sqg} has the dimensionality of (length)$^2$, so that a signal of spacetime fuzziness would be detectable if
\[
\left(\frac{\ell_*}{L}\right)^{2(1-\a)} = \frac{\s_{\rm QG}^2}{L^2} \sim \s^2_{\rm exp} \simeq f\,\cS^2(f)\big|_{f=\frac{1}{L}}\,,
\]
leading to
\be\label{gS}
\cS(f)= \left(\frac{1}{\ell_*}\right)^{\a-1} f^{\frac12-\a}\qquad \Longrightarrow \qquad \a=\frac{\ln\left(\frac{\cS}{\ell_*\sqrt{f}}\right)}{\ln\left(\frac{1}{\ell_* f}\right)}\,.
\ee

%%%%%%%%%%%%%%%%%%%%%%%%%%%%%%%%%%%%%%%%%%%%%%%%%%%%%%%%%%%%%%%%%%%%%%%%%%%%%%%%%

\subsection{Results in quantum gravity}

In the worst-case scenario where $\ell_*$ is of Planckian size, one might believe it impossible to probe with an instrument of a macroscopic size $L$ of order of the kilometer (or millions of kilometers, in the case of LISA). However, $L$ does not appear in \Eq{gS} and, if $\a$ is small enough, the detector may even catch the stochastic background from spacetime fuzziness. Setting $\ell_* =\lp$ in \Eq{gS}, we get an upper bound on $\a$ for all the main interferometers in operation, under construction, or proposed for the near future, ranging from $\a<0.47$ for Ligo-Virgo-KAGRA and DECIGO to $\a<0.54$ for LISA \cite{Calcagni:2019kzo}. This translates into a bound on the small-scale Hausdorff dimension of spacetime,
\be\label{dhuvbou}
\dh^{\rm UV}<1.9\,,%\dh^{\rm UV}=\ds^{\rm UV}<2.9\,.
\ee
very close to the value found in certain kinematical states of LQG/SF/GFT. Note that an infinitely sensitive instrument not detecting quantum-gravity noise would push the bound to $\a\sim 0$, which would mean that there is no dimensional flow ($\dh=4$), i.e., the condition of application of the flow-equation theorem does not hold. Therefore, one could not interpret this result as having a spacetime with zero UV dimension. In particular, the Hausdorff dimension in non-local quantum gravity does not change and we do not expect any imprint in the strain noise. Any constraint from \Eq{gS} would be physically helpful only in the case where a strain noise of quantum-gravity origin were actually detected.

%%%%%%%%%%%%%%%%%%%%%%%%%%%%%%%%%%%%%%%%%%%%%%%%%%%%%%%%%%%%%%%%%%%%%%%%%%%%%%%%%
%%%%%%%%%%%%%%%%%%%%%%%%%%%%%%%%%%%%%%%%%%%%%%%%%%%%%%%%%%%%%%%%%%%%%%%%%%%%%%%%%

\section{Conclusions}

Table \ref{tab1} summarizes the cosmological models of quantum gravity that we have explored here in relation with GW physics. To the model discussed in this review, we add pre-big-bang cosmology, where $r<0.01$ at the pivot scale $k_0=0.05\,{\rm Mpc}^{-1}$ and the stochastic GW background is blue-tilted and can reach the sensitivity of present and future interferometers \cite{Gasperini:2016gre}. As one can appreciate, their great majority does not give rise to any observable signal, as expected on the grounds that Planck-scale corrections do not have an impact on the production and propagation of gravitational waves. Still, in some cases there is margin for moderate hope to see something, or at least to place meaningful constraints, in the near future. All these results would deserve further critical scrutiny because they are as good as the assumptions made to obtain them. For example, it is not yet clear whether the bump in the spectral dimension found in kinematical quantum states in LQG/SF/GFT and giving rise to a potential detectable deviation from the luminosity distance \cite{Calcagni:2019kzo} is an artifact or a physical feature of the model \cite{COT3}.
\begin{table}
\centering
\begin{tabular}{l|p{1cm}p{1cm}p{1cm}p{1.5cm}}\hline
																				  & $\Om_\textsc{gw}$	&	$c_\textsc{gw}$	& $d_L$     & Strain noise \\\hline\hline
Stelle gravity  													& 									&  		      	 	  &  			    & 						 \\
String theory (low-energy limit)					& 									&           	    &  				  & 						 \\
Asymptotic safety													& 									&           	    &  			   	& ?						 \\
LQG/SF/GFT        										 		&  			      			&           	    & \ding{51} & \ding{51}    \\
{\small\qquad loop quantum cosmology}		  &  									&                 &  			    & 						 \\
Causal dynamical triangulations (phase C) & ?								  &  			      	  &  			    & ?						 \\
Non-local quantum gravity                 & 									& 		     	      &           &              \\
Ho\v{r}ava--Lifshitz gravity							&  			        	  & \ding{51}  	    &  				 	& 						 \\\hline
Pre-big-bang cosmology									  & \ding{51}     	  &  		  	        & 				  & 				 		 \\
String-gas cosmology										  & \ding{51}     	  &  		  	        & 				  & 				 		 \\
New ekpyrotic scenario									  & \ding{51}      	  &  		  	        & 				  & 				 		 \\
Brandenberger--Ho non-commutative inflation& \ding{51}        &  	    	        & 				  & 						 \\
$\k$-Minkowski spacetime									& ?								  &    	            & 				  & ?            \\
Padmanabhan's non-local model							&  			        	  &  			      	  &  				 	& 						 \\
Multi-fractional spacetimes               & \ding{51}		      & \ding{51}       & \ding{51} & \ding{51}    \\\hline\hline
\end{tabular}
\caption{\label{tab1} Observability in GW data of various theories of quantum gravity using the stochastic GW background ($\Om_\textsc{gw}$), the propagation speed of gravitons ($c_\textsc{gw}$), the luminosity distance ($d_L$) and the strain noise. A tick indicates that the theory might give a detectable signal, empty cells correspond to theories which cannot produce such a signal, and question marks are placed where no prediction has been calculated so far.}
\end{table}

The number of ticks in the table is not related to the quality of science one can do with each theory. Some models have many ticks because they can be tuned more easily than others, while some have no tick at all because their predictions are rigorous but consistently below detection threshold. Also, the table is by no means exhaustive. While we have spent some time on the propagation of GWs, we have not discussed the constraints that can come from the production of GWs at inspiral and merger phases \cite{Yunes:2016jcc,BYY} or from the horizon structure of merging black holes \cite{AMY}, topics that would require another review and more advances from the theoretical side than those currently available for the theories in our list, list that, as we already stressed, is not comprehensive.

The main message to take on board from this review is that, in certain models of quantum gravity, short-scale modifications can leave an imprint on GWs when they are accumulated on cosmic distances or amplified by the cosmic expansion via mechanisms that go beyond effective field theory. By the first half of this century, GW interferometers should be able to give a deeper insight into the physics of quantum gravity.

%%%%%%%%%%%%%%%%%%%%%%%%%%%%%%%%%%%%%%%%%%%%%%%%%%%%%%%%%%%%%%%%%%%%%%%%%%%%%%%%%%%%%%%%%
%%%%%%%%%%%%%%%%%%%%%%%%%%%%%%%%%%%%%%%%%%%%%%%%%%%%%%%%%%%%%%%%%%%%%%%%%%%%%%%%%%%%%%%%%

\section*{Acknowledgments}

\noindent The author is supported by the I+D grant FIS2017-86497-C2-2-P of the Spanish Ministry of Science and Innovation and acknowledges networking support by the COST Action CA18108.

%%%%%%%%%%%%%%%%%%%%%%%%%%%%%%%%%%%%%%%%%%%%%%%%%%%%%%%%%%%%%%%%%%%%%%%%%%%%%%%%%
%%%%%%%%%%%%%%%%%%%%%%%%%%%%%%%%%%%%%%%%%%%%%%%%%%%%%%%%%%%%%%%%%%%%%%%%%%%%%%%%%

%%%%%%%%%%%%%%%%%%%%%%%% reference.tex %%%%%%%%%%%%%%%%%%%%%
% sample references
% 
% Use this file as a template for your own input.
%
%%%%%%%%%%%%%%%%%%%%%%%% Springer%%%%%%%%%%%%%%%%%%%%%%%%%%
%
% BibTeX users please use
% \bibliographystyle{}
% \bibliography{}

\begin{thebibliography}{99.}
\bibitem{TheLIGOScientific:2016src} \au{BP}{Abbott} {et al.} [LIGO Scientific and \textsc{Virgo} Collaborations], \tia{Tests of general relativity with GW150914} \doinn{10.1103/PhysRevLett.116.221101}{Phys.\ Rev.\ Lett.}{116}{221101}{2016}; \doinn{10.1103/PhysRevLett.121.129902}{Erratum-ibid.}{121}{129902}{2018} [\arX{1602.03841}].
\bibitem{LIGOScientific:2020stg} LIGO Scientific Collaboration and \textsc{Virgo} Collaboration, \tia{GW190412: observation of a binary-black-hole coalescence with asymmetric masses} \doin{10.1103/PhysRevD.102.043015}{Phys.\ Rev.}{D}{102}{043015}{2020} [\arX{2004.08342}].
\bibitem{Ame98} \au{G}{Amelino-Camelia}, \tia{An interferometric gravitational wave detector as a quantum gravity apparatus} \doinn{10.1038/18377}{Nature}{398}{216}{1998} [\oarX{gr-qc/9808029}].
\bibitem{NgDa2} \au{YJ}{Ng}, \au{H}{Van Dam}, \tia{Measuring the foaminess of space-time with gravity-wave interferometers} \doinn{10.1023/A:1003745212871}{Found.\ Phys.}{30}{795}{2000} [\oarX{gr-qc/9906003}].
\bibitem{Ame13} \au{G}{Amelino-Camelia}, \tia{Quantum-spacetime phenomenology} \doinn{10.12942/lrr-2013-5}{Living Rev.\ Rel.}{16}{5}{2013} [\arX{0806.0339}].
\bibitem{EMNan} \au{J}{Ellis}, \au{NE}{Mavromatos}, \au{DV}{Nanopoulos}, \tia{Comments on graviton propagation in light of GW150914} \doin{10.1142/S0217732316750018}{Mod.\ Phys.\ Lett.}{A}{31}{1650155}{2016} [\arX{1602.04764}].
%\bibitem{qGW}   \au{G}{Calcagni}, \tia{Lorentz violations in multifractal spacetimes} \doin{10.1140/epjc/s10052-017-4841-6}{Eur.\ Phys.\ J.}{C}{77}{291}{2017} [\arX{1603.03046}].
\bibitem{Yunes:2016jcc}   \au{N}{Yunes}, \au{K}{Yagi}, \au{F}{Pretorius}, \tia{Theoretical physics implications of the binary black-hole merger GW150914} \doin{10.1103/PhysRevD.94.084002}{Phys.\ Rev.}{D}{94}{084002}{2016} [\arX{1603.08955}].
\bibitem{ArCa2} \au{M}{Arzano}, \au{G}{Calcagni}, \tia{What gravity waves are telling about quantum spacetime} \doin{10.1103/PhysRevD.93.124065}{Phys.\ Rev.}{D}{93}{124065}{2016} [\arX{1604.00541}].
\bibitem{Kobakhidze:2016cqh} \au{A}{Kobakhidze}, \au{C}{Lagger}, \au{A}{Manning}, \tia{Constraining noncommutative spacetime from GW150914} \doin{10.1103/PhysRevD.94.064033}{Phys.\ Rev.}{D}{94}{064033}{2016} [\arX{1607.03776}].
\bibitem{revmu} \au{G}{Calcagni}, \tia{Multifractional theories: an unconventional review} \doij{10.1007/JHEP03(2017)138}{J.\ High Energy Phys.}{1703}{138}{2017} [\arX{1612.05632}].
\bibitem{ACCR}  \au{G}{Amelino-Camelia}, \au{G}{Calcagni}, \au{M}{Ronco}, \tia{Imprint of quantum gravity in the dimension and fabric of spacetime} \doin{10.1016/j.physletb.2017.10.032}{Phys.\ Lett.}{B}{774}{630}{2017} [\arX{1705.04876}].
\bibitem{BYY}   \au{E}{Berti}, \au{K}{Yagi}, \au{N}{Yunes}, \tia{Extreme gravity tests with gravitational waves from compact binary coalescences: (I) inspiral-merger} \doinn{10.1007/s10714-018-2362-8}{Gen.\ Rel.\ Grav.}{50}{46}{2018} [\arX{1801.03208}].
\bibitem{Bosso:2018ckz} \au{P}{Bosso}, \au{S}{Das}, \au{RB}{Mann}, \tia{Potential tests of the generalized uncertainty principle in the advanced LIGO experiment} \doin{10.1016/j.physletb.2018.08.061}{Phys.\ Lett.}{B}{785}{498}{2018} [\arX{1804.03620}].
\bibitem{TaYa}  \au{S}{Tahura}, \au{K}{Yagi}, \tia{Parametrized post-Einsteinian gravitational waveforms in various modified theories of gravity} \doin{10.1103/PhysRevD.98.084042}{Phys.\ Rev.}{D}{98}{084042}{2018} [\arX{1809.00259}].
\bibitem{AMY}   \au{A}{Addazi}, \au{A}{Marcian\`o}, \au{N}{Yunes}, \tia{Can we probe Planckian corrections at the horizon scale with gravitational waves?} \doinn{10.1103/PhysRevLett.122.081301}{Phys.\ Rev.\ Lett.}{122}{081301}{2019} [\arX{1810.10417}].
\bibitem{Mas18} \au{A}{Maselli}, \au{P}{Pani}, \au{V}{Cardoso}, \au{T}{Abdelsalhin}, \au{L}{Gualtieri}, \au{V}{Ferrari}, \tia{From micro to macro and back: probing near-horizon quantum structures with gravitational waves} \doinn{10.1088/1361-6382/ab30ff}{Class.\ Quantum Grav.}{36}{167001}{2019} [\arX{1811.03689}].
\bibitem{Calcagni:2019kzo} \au{G}{Calcagni}, \au{S}{Kuroyanagi}, \au{S}{Marsat}, \au{M}{Sakellariadou}, \au{N}{Tamanini}, \au{G}{Tasinato}, \tia{Gravitational-wave luminosity distance in quantum gravity} \doin{10.1016/j.physletb.2019.135000}{Phys.\ Lett.}{B}{798}{135000}{2019} [\arX{1904.00384}].
\bibitem{Giddings:2019ujs} \au{S.B}{Giddings}, \au{S}{Koren}, \au{G}{Trevi\~no}, \tia{Exploring strong-field deviations from general relativity via gravitational waves} \doin{10.1103/PhysRevD.100.044005}{Phys.\ Rev.}{D}{100}{044005}{2019} [\arX{1904.04258}].
\bibitem{Belgacem:2019pkk} \au{E}{Belgacem} {\it et al.} [LISA Cosmology Working Group], \tia{Testing modified gravity at cosmological distances with LISA standard sirens} \doij{10.1088/1475-7516/2019/07/024}{JCAP}{07}{024}{2019} [\arX{1906.01593}].
\bibitem{Calcagni:2019ngc} \au{G}{Calcagni}, \au{S}{Kuroyanagi}, \au{S}{Marsat}, \au{M}{Sakellariadou}, \au{N}{Tamanini}, \au{G}{Tasinato}, \tia{Quantum gravity and gravitational-wave astronomy} \doij{10.1088/1475-7516/2019/10/012}{JCAP}{10}{012}{2019} [\arX{1907.02489}].
\bibitem{Wang:2020pgu} \au{S}{Wang}, \au{Z.-C}{Zhao}, \tia{Tests of CPT invariance in gravitational waves with LIGO-Virgo catalog GWTC-1} \doin{10.1140/epjc/s10052-020-08628-x}{Eur.\ Phys.\ J.}{C}{80}{1032}{2020} [\arX{2002.00396}].
\bibitem{Garcia-Chung:2020zyq} \au{A}{Garcia-Chung}, \au{J.B}{Mertens}, \au{S}{Rastgoo}, \au{Y}{Tavakoli}, \au{P}{Vargas Moniz}, \tia{Propagation of quantum gravity-modified gravitational waves on a classical FLRW spacetime} \arX{2012.09366}.
\bibitem{For82} \au{LH}{Ford}, \tia{Gravitational radiation by quantum systems} \doinn{10.1016/0003-4916(82)90115-4}{Ann.\ Phys.\ (N.Y.)}{144}{238}{1982}.
\bibitem{Wal84} \au{RM}{Wald}, \book{General Relativity}{The University of Chicago Press}{Chicago}{IL}{1984}.
\bibitem{PaG81} \au{DN}{Page}, \au{CD}{Geilker}, \tia{Indirect evidence for quantum gravity} \doinn{10.1103/PhysRevLett.47.979}{Phys.\ Rev.\ Lett.}{47}{979}{1981}.
\bibitem{Car08} \au{S}{Carlip}, \tia{Is quantum gravity necessary?} \doinn{10.1088/0264-9381/25/15/154010}{Class.\ Quantum Grav.}{25}{154010}{2008} [\arX{0803.3456}].
\bibitem{Bartolo:2016ami} \au{N}{Bartolo} {et al.}, \tia{Science with the space-based interferometer LISA. IV: Probing inflation with gravitational waves} \doij{10.1088/1475-7516/2016/12/026}{JCAP}{12}{026}{2016} [\arX{1610.06481}].
\bibitem{Kuroyanagi:2018csn} \au{S}{Kuroyanagi}, \au{T}{Chiba}, \au{T}{Takahashi}, \tia{Probing the universe through the stochastic gravitational wave background} \doij{10.1088/1475-7516/2018/11/038}{JCAP}{11}{038}{2018} [\arX{1807.00786}].
\bibitem{Ori09} \au{D}{Oriti} ed., \book{Approaches to Quantum Gravity}{Cambridge University Press}{Cambridge}{UK}{2009}.
\bibitem{Fousp} \au{GFR}{Ellis}, \au{J}{Murugan}, \au{A}{Weltman} eds., \book{Foundations of Space and Time}{Cambridge University Press}{Cambridge}{UK}{2012}.
\bibitem{CQC} \au{G}{Calcagni}, \books{\href{10.1007/978-3-319-41127-9}{\cob Classical and Quantum Cosmology}}{Springer}{Switzerland}{2017}.
\bibitem{Ste77} \au{KS}{Stelle}, \tia{Renormalization of higher-derivative quantum gravity} \doin{10.1103/PhysRevD.16.953}{Phys.\ Rev.}{D}{16}{953}{1977}.
\bibitem{Ste78} \au{KS}{Stelle}, \tia{Classical gravity with higher derivatives} \doinn{10.1007/BF00760427}{Gen.\ Rel.\ Grav.}{9}{353}{1978}.
\bibitem{ALS}   \au{M}{Asorey}, \au{JL}{L\'opez}, \au{IL}{Shapiro}, \tia{Some remarks on high derivative quantum gravity} \doin{10.1142/S0217751X97002991}{Int.\ J.\ Mod.\ Phys.}{A}{12}{5711}{1997} [\oarX{hep-th/9610006}].
\bibitem{AAM}   \au{A}{Accioly}, \au{A}{Azeredo}, \au{H}{Mukai}, \tia{Propagator, tree-level unitarity and effective nonrelativistic potential for higher-derivative gravity theories in $D$ dimensions} \doinn{10.1063/1.1415743}{J.\ Math.\ Phys.\ (N.Y.)}{43}{473}{2002}.
\bibitem{Pol98} \au{J}{Polchinski}, \book{String Theory}{Cambridge University Press}{Cambridge}{UK}{1998}.
\bibitem{BBSb}  \au{K}{Becker}, \au{M}{Becker}, \au{JH}{Schwarz}, \book{String Theory and M-Theory}{Cambridge University Press}{Cambridge}{U.K.}{2007}.
\bibitem{Zwi09} \au{B}{Zwiebach}, \book{A First Course in String Theory}{Cambridge University Press}{Cambridge}{U.K.}{2009}.
\bibitem{Baumann:2014nda} \au{D}{Baumann}, \au{L}{McAllister}, \book{\href{http://dx.doi.org/10.1017/CBO9781316105733}{\cob Inflation and String Theory}}{Cambridge University Press}{Cambridge}{U.K.}{2015} [\arX{1404.2601}].
\bibitem{Wei79} \au{S}{Weinberg}, \tia{Ultraviolet divergences in quantum gravity} in \au{SW}{Hawking}, \au{W}{Israel} eds., \book{General Relativity: An Einstein Centenary Survey}{Cambridge University Press}{Cambridge}{UK}{1979}. %pp.\ 790-831
\bibitem{Reu1}  \au{M}{Reuter}, \tia{Nonperturbative evolution equation for quantum gravity} \doin{10.1103/PhysRevD.57.971}{Phys.\ Rev.}{D}{57}{971}{1998} [\oarX{hep-th/9605030}].
\bibitem{NiR}   \au{M}{Niedermaier}, \au{M}{Reuter}, \tia{The asymptotic safety scenario in quantum gravity} \doinn{10.12942/lrr-2006-5}{Living Rev.\ Rel.}{9}{5}{2006}.
\bibitem{Nie06} \au{M}{Niedermaier}, \tia{The asymptotic safety scenario in quantum gravity: an introduction} \doinn{10.1088/0264-9381/24/18/R01}{Class.\ Quantum Grav.}{24}{R171}{2007} [\oarX{gr-qc/0610018}].
\bibitem{CPR}   \au{A}{Codello}, \au{R.\ Percacci}, \au{C}{Rahmede}, \tia{Investigating the ultraviolet properties of gravity with a Wilsonian renormalization group equation} \doinn{10.1016/j.aop.2008.08.008}{Annals Phys.}{324}{414}{2009} [\arX{0805.2909}].
\bibitem{Lit11} \au{DF}{Litim}, \tia{Renormalisation group and the Planck scale} \doin{10.1098/rsta.2011.0103}{Phil.\ Trans.\ Roy.\ Soc.\ Lond.}{A}{369}{2759}{2011} [\arX{1102.4624}].
\bibitem{RSnax} \au{M}{Reuter}, {F}{Saueressig}, \tia{Asymptotic safety, fractals, and cosmology} \doinn{10.1007/978-3-642-33036-0_8}{Lect.\ Notes Phys.}{863}{185}{2013} [\arX{1205.5431}].
\bibitem{rov07} \au{C}{Rovelli}, \book{Quantum Gravity}{Cambridge University Press}{Cambridge}{UK}{2007}.
\bibitem{thi01} \au{T}{Thiemann}, \book{Modern Canonical Quantum General Relativity}{Cambridge University Press}{Cambridge}{UK}{2007}; \tia{Introduction to modern canonical quantum general relativity} \oarX{gr-qc/0110034}.
\bibitem{Per03} \au{A}{Perez}, \tia{Spin foam models for quantum gravity} \doinn{10.1088/0264-9381/20/6/202}{Class.\ Quantum Grav.}{20}{R43}{2003} [\oarX{gr-qc/0301113}].
\bibitem{Rov10} \au{C}{Rovelli}, \tia{A new look at loop quantum gravity} \doinn{10.1088/0264-9381/28/11/114005}{Class.\ Quantum Grav.}{28}{114005}{2011} [\arX{1004.1780}].
\bibitem{Per13} \au{A}{Perez}, \tia{The spin-foam approach to quantum gravity} \doinn{10.12942/lrr-2013-3}{Living Rev.\ Rel.}{16}{3}{2013}.% [\arX{1205.2019}]
\bibitem{Fre05} \au{L}{Freidel}, \tia{Group field theory: an overview} \doinn{10.1007/s10773-005-8894-1}{Int.\ J.\ Theor.\ Phys.}{44}{1769}{2005} [\oarX{hep-th/0505016}].
\bibitem{BaO11} \au{A}{Baratin}, \au{D}{Oriti}, \tia{Ten questions on group field theory (and their tentative answers)} \doinn{10.1088/1742-6596/360/1/012002}{J.\ Phys.\ Conf.\ Ser.}{360}{012002}{2012} [\arX{1112.3270}].
\bibitem{Ori13} \au{D}{Oriti}, \tia{Group field theory as the second quantization of loop quantum gravity} \doinn{10.1088/0264-9381/33/8/085005}{Class.\ Quantum Grav.}{33}{085005}{2016} [\arX{1310.7786}].
\bibitem{GiSi}  \au{S}{Gielen}, \au{L}{Sindoni}, \tia{Quantum cosmology from group field theory condensates: a review} \doinn{10.3842/SIGMA.2016.082}{SIGMA}{12}{082}{2016} [\arX{1602.08104}].
\bibitem{AmJ}   \au{J}{Ambj{\o}rn}, \au{J}{Jurkiewicz}, \tia{Scaling in four-dimensional quantum gravity} \doin{10.1016/0550-3213(95)00303-A}{Nucl.\ Phys.}{B}{451}{643}{1995} [\oarX{hep-th/9503006}].
\bibitem{AJL4}  \au{J}{Ambj{\o}rn}, \au{J}{Jurkiewicz}, \au{R}{Loll}, \tia{Spectral dimension of the universe} \doinn{10.1103/PhysRevLett.95.171301}{Phys.\ Rev.\ Lett.}{95}{171301}{2005} [\oarX{hep-th/0505113}].
\bibitem{AJL5}  \au{J}{Ambj{\o}rn}, \au{J}{Jurkiewicz}, \au{R}{Loll}, \tia{Reconstructing the universe} \doin{10.1103/PhysRevD.72.064014}{Phys.\ Rev.}{D}{72}{064014}{2005} [\oarX{hep-th/0505154}].
\bibitem{lol08} \au{R}{Loll}, \tia{The emergence of spacetime, or, quantum gravity on your desktop} \doinn{10.1088/0264-9381/25/11/114006}{Class.\ Quantum Grav.}{25}{114006}{2008} [\arX{0711.0273}].
\bibitem{AJL8}  \au{J}{Ambj{\o}rn}, \au{J}{Jurkiewicz}, \au{R}{Loll}, \tia{Causal dynamical triangulations and the quest for quantum gravity} in ref.\ \cite{Fousp} [\arX{1004.0352}].
\bibitem{AGJL4} \au{J}{Ambj{\o}rn}, \au{A}{G\"orlich}, \au{J}{Jurkiewicz}, \au{R}{Loll}, \tia{Nonperturbative quantum gravity} \doinn{10.1016/j.physrep.2012.03.007}{Phys.\ Rept.}{519}{127}{2012} [\arX{1203.3591}].
\bibitem{CoJu}  \au{DN}{Coumbe}, \au{J}{Jurkiewicz}, \tia{Evidence for asymptotic safety from dimensional reduction in causal dynamical triangulations} \doij{10.1007/JHEP03(2015)151}{JHEP}{03}{151}{2015} [\arX{1411.7712}].
\bibitem{CoDo}  \au{JH}{Cooperman}, \au{M}{Dorghabekov}, \tia{Setting the physical scale of dimensional reduction in causal dynamical triangulations} \doin{10.1103/PhysRevD.100.026014}{Phys.\ Rev.}{D}{100}{026014}{2019} [\arX{1812.09331}].
\bibitem{Kuz89} \au{YuV}{Kuz'min}, \tia{The convergent nonlocal gravitation} Sov.\ J.\ Nucl.\ Phys.\ {\bf 50}, 1011 (1989) [Yad.\ Fiz.\ {\bf 50}, 1630 (1989)].
\bibitem{Tom97} \au{ET}{Tomboulis}, \tia{Super-renormalizable gauge and gravitational theories} \oarX{hep-th/9702146}.
\bibitem{Modesto:2011kw} \au{L}{Modesto}, \tia{Super-renormalizable quantum gravity} \doin{10.1103/PhysRevD.86.044005}{Phys.\ Rev.}{D}{86}{044005}{2012} [\arX{1107.2403}].
\bibitem{BGKM}  \au{T}{Biswas}, \au{E}{Gerwick}, \au{T}{Koivisto}, \au{A}{Mazumdar}, \tia{Towards singularity and ghost free theories of gravity} \doinn{10.1103/PhysRevLett.108.031101}{Phys.\ Rev.\ Lett.}{108}{031101}{2012} [\arX{1110.5249}].
\bibitem{Modesto:2017sdr} \au{L}{Modesto}, \au{L}{Rachwa\l}, \tia{Nonlocal quantum gravity: a review} \doin{10.1142/S0218271817300208}{Int.\ J.\ Mod.\ Phys.}{D}{26}{1730020}{2017}.
\bibitem{BrCM} \au{F}{Briscese}, \au{G}{Calcagni}, \au{L}{Modesto}, \tia{Nonlinear stability in nonlocal gravity} \doin{10.1103/PhysRevD.99.084041}{Phys.\ Rev.}{D}{99}{084041}{2019} [\arX{1901.03267}].
\bibitem{Hor09} \au{P}{Ho\v{r}ava}, \tia{Quantum gravity at a Lifshitz point} \doin{10.1103/PhysRevD.79.084008}{Phys.\ Rev.}{D}{79}{084008}{2009} [\arX{0901.3775}].
\bibitem{Hor3}  \au{P}{Ho\v{r}ava}, \tia{Spectral dimension of the universe in quantum gravity at a Lifshitz point} \doinn{10.1103/PhysRevLett.102.161301}{Phys.\ Rev.\ Lett.}{102}{161301}{2009} [\arX{0902.3657}].
\bibitem{HoMe}  \au{P}{Ho\v{r}ava}, \au{CM}{Melby-Thompson}, \tia{General covariance in quantum gravity at a Lifshitz point} \doin{10.1103/PhysRevD.82.064027}{Phys.\ Rev.}{D}{82}{064027}{2010} [\arX{1007.2410}].
\bibitem{Gasperini:1992em} \au{M}{Gasperini}, \au{G}{Veneziano}, \tia{Pre-big bang in string cosmology} \doinn{10.1016/0927-6505(93)90017-8}{Astropart.\ Phys.}{1}{317}{1993} [\arX{hep-th/9211021}].
\bibitem{Bra11} \au{RH}{Brandenberger}, \tia{String gas cosmology: progress and problems} \doinn{10.1088/0264-9381/28/20/204005}{Class.\ Quantum Grav.}{28}{204005}{2011} [\arX{1105.3247}].
\bibitem{Brandenberger:2015kga} \au{RH}{Brandenberger}, \tia{String gas cosmology after Planck} \doinn{10.1088/0264-9381/32/23/234002}{Class.\ Quantum Grav.}{32}{234002}{2015} [\arX{1505.02381}].
\bibitem{Brandenberger:2020tcr} \au{R}{Brandenberger}, \au{Z}{Wang}, \tia{Nonsingular ekpyrotic cosmology with a nearly scale-invariant spectrum of cosmological perturbations and gravitational waves} \doin{10.1103/PhysRevD.101.063522}{Phys.\ Rev.}{D}{101}{063522}{2020} [\arX{2001.00638}].
\bibitem{Brandenberger:2020eyf} \au{R}{Brandenberger}, \au{Z}{Wang}, \tia{Ekpyrotic cosmology with a zero-shear S-brane} \doin{10.1103/PhysRevD.102.023516}{Phys.\ Rev.}{D}{102}{023516}{2020} [\arX{2004.06437}].
\bibitem{BH}    \au{R}{Brandenberger}, \au{P-M}{Ho}, \tia{Noncommutative spacetime, stringy spacetime uncertainty principle, and density fluctuations} \doin{10.1103/PhysRevD.66.023517}{Phys.\ Rev.}{D}{66}{023517}{2002} [\oarX{hep-th/0203119}].
\bibitem{Calcagni:2013lya} \au{G}{Calcagni}, \au{S}{Kuroyanagi}, \au{J}{Ohashi}, \au{S}{Tsujikawa}, \tia{Strong Planck constraints on braneworld and non-commutative inflation} \doij{10.1088/1475-7516/2014/03/052}{JCAP}{1403}{052}{2014} [\arX{1310.5186}].
\bibitem{Sza01} \au{RJ}{Szabo}, \tia{Quantum field theory on noncommutative spaces} \doinn{10.1016/S0370-1573(03)00059-0}{Phys.\ Rept.}{378}{207}{2003} [\oarX{hep-th/0109162}].
\bibitem{ADKLW} \au{P}{Aschieri}, \au{M}{Dimitrijevic}, \au{P}{Kulish}, \au{F}{Lizzi}, \au{J}{Wess}, \book{Noncommutative Spacetimes}{Springer}{Berlin}{Germany}{2009}.
\bibitem{Ben08} \au{D}{Benedetti}, \tia{Fractal properties of quantum spacetime} \doinn{10.1103/PhysRevLett.102.111303}{Phys.\ Rev.\ Lett.}{102}{111303}{2009} [\arX{0811.1396}].
\bibitem{ArTr}  \au{M}{Arzano}, \au{T}{Trze\'sniewski}, \tia{Diffusion on $\kappa$-Minkowski space} \doin{10.1103/PhysRevD.89.124024}{Phys.\ Rev.}{D}{89}{124024}{2014} [\arX{1404.4762}].
\bibitem{Eckstein:2020gjd} \au{M}{Eckstein} and \au{T}{Trze\'sniewski}, \tia{Spectral dimensions and dimension spectra of quantum spacetimes}
\doin{10.1103/PhysRevD.102.086003}{Phys.\ Rev.}{D}{102}{086003}{2020} [\arX{2005.14210}].
\bibitem{Pad98} \au{T}{Padmanabhan}, \tia{Quantum structure of space-time and black hole entropy} \doinn{10.1103/PhysRevLett.81.4297}{Phys.\ Rev.\ Lett.}{81}{4297}{1998} [\oarX{hep-th/9801015}].
\bibitem{Pad99} \au{T}{Padmanabhan}, \tia{Event horizon: magnifying glass for Planck length physics} \doin{10.1103/PhysRevD.59.124012}{Phys.\ Rev.}{D}{59}{124012}{1999} [\oarX{hep-th/9801138}].
\bibitem{ArCa1} \au{M}{Arzano}, \au{G}{Calcagni}, \tia{Black-hole entropy and minimal diffusion} \doin{10.1103/PhysRevD.88.084017}{Phys.\ Rev.}{D}{88}{084017}{2013} [\arX{1307.6122}].
\bibitem{Akrami:2018odb} \au{Y}{Akrami} {et al.} [Planck Collaboration], \tia{Planck 2018 results. X. Constraints on inflation} \doinn{10.1051/0004-6361/201833887}{Astron.\ Astrophys.}{641}{A10}{2020} [\arX{1807.06211}].
\bibitem{Kuroyanagi:2014nba} \au{S}{Kuroyanagi}, \au{T}{Takahashi} and \au{S}{Yokoyama}, \tia{Blue-tilted tensor spectrum and thermal history of the universe} \doij{10.1088/1475-7516/2015/02/003}{JCAP}{1502}{003}{2015} [\arX{1407.4785}].
%\bibitem{Calcagni:2017amr} \au{G}{Calcagni}, \tia{Multifractional spacetimes from the Standard Model to cosmology} \doinn{10.1142/S0219887819400048}{Int.\ J.\ Geom.\ Meth.\ Mod.\ Phys.}{16}{1940004}{2018} [\arX{1709.07844}].
\bibitem{TheLIGOScientific:2016wyq} \au{BP}{Abbott} {et al.} [LIGO Scientific and \textsc{Virgo} Collaborations], \tia{GW150914: implications for the stochastic gravitational wave background from binary black holes} \doinn{10.1103/PhysRevLett.116.131102}{Phys.\ Rev.\ Lett.}{116}{131102}{2016} [\arX{1602.03847}].
\bibitem{Abbott:2017xzg} \au{BP}{Abbott} {et al.} [LIGO Scientific and \textsc{Virgo} Collaborations], \tia{GW170817: implications for the stochastic gravitational-wave background from compact binary coalescences} \doinn{10.1103/PhysRevLett.120.091101}{Phys.\ Rev.\ Lett.}{120}{091101}{2018} [\arX{1710.05837}].
\bibitem{Akutsu:2018axf} \au{T}{Akutsu} {et al.} [KAGRA], \tia{KAGRA: 2.5 generation interferometric gravitational wave detector} \doinn{10.1038/s41550-018-0658-y}{Nat.\ Astron.}{3}{35}{2019} [\arX{1811.08079}].
\bibitem{Caprini:2019pxz} \au{C}{Caprini}, \au{DG}{Figueroa}, \au{R}{Flauger}, \au{G}{Nardini}, \au{M}{Peloso}, \au{M}{Pieroni}, \au{A}{Ricciardone}, \au{G}{Tasinato}, \tia{Reconstructing the spectral shape of a stochastic gravitational wave background with LISA} \doij{10.1088/1475-7516/2019/11/017}{JCAP}{1911}{017}{2019} [\arX{1906.09244}].
\bibitem{Maggiore:2019uih} \au{M}{Maggiore}, \au{C}{Van Den Broeck}, \au{N}{Bartolo}, \au{E}{Belgacem}, \au{D}{Bertacca}, \au{MA}{Bizouard}, \au{M}{Branchesi}, \au{S}{Clesse}, \au{S}{Foffa}, \au{J}{Garc\'ia-Bellido}, \au{S}{Grimm}, \au{J}{Harms}, \au{T}{Hinderer}, \au{S}{Matarrese}, \au{C}{Palomba}, \au{M}{Peloso}, \au{A}{Ricciardone}, \au{M}{Sakellariadou}, \tia{Science case for the Einstein Telescope} \doij{10.1088/1475-7516/2020/03/050}{JCAP}{03}{050}{2020} [\arX{1912.02622}].
\bibitem{Seto:2001qf} \au{N}{Seto}, \au{S}{Kawamura}, \au{T}{Nakamura}, \tia{Possibility of direct measurement of the acceleration of the universe using 0.1-Hz band laser interferometer gravitational wave antenna in space} \doinn{10.1103/PhysRevLett.87.221103}{Phys.\ Rev.\ Lett.}{87}{221103}{2001} [\oarX{astro-ph/0108011}].
\bibitem{Kawamura:2011zz} \au{S}{Kawamura} {et al.}, \tia{The Japanese space gravitational wave antenna: DECIGO} \doinn{10.1088/0264-9381/28/9/094011}{Class.\ Quant.\ Grav.}{28}{094011}{2011}.
\bibitem{Kawamura:2020pcg} \au{S}{Kawamura} {et al.}, %M.~Ando, N.~Seto, S.~Sato, M.~Musha, I.~Kawano, J.~Yokoyama, T.~Tanaka, K.~Ioka, T.~Akutsu, T.~Takashima, K.~Agatsuma, A.~Araya, N.~Aritomi, H.~Asada, T.~Chiba, S.~Eguchi, M.~Enoki, M.~K.~Fujimoto, R.~Fujita, T.~Futamase, T.~Harada, K.~Hayama, Y.~Himemoto, T.~Hiramatsu, F.~L.~Hong, M.~Hosokawa, K.~Ichiki, S.~Ikari, H.~Ishihara, T.~Ishikawa, Y.~Itoh, T.~Ito, S.~Iwaguchi, K.~Izumi, N.~Kanda, S.~Kanemura, F.~Kawazoe, S.~Kobayashi, K.~Kohri, Y.~Kojima, K.~Kokeyama, K.~Kotake, S.~Kuroyanagi, K.~i.~Maeda, S.~Matsushita, Y.~Michimura, T.~Morimoto, S.~Mukohyama, K.~Nagano, S.~Nagano, T.~Naito, K.~Nakamura, T.~Nakamura, H.~Nakano, K.~Nakao, S.~Nakasuka, Y.~Nakayama, K.~Nakazawa, A.~Nishizawa, M.~Ohkawa, K.~Oohara, N.~Sago, M.~Saijo, M.~Sakagami, S.~i.~Sakai, T.~Sato, M.~Shibata, H.~Shinkai, A.~Shoda, K.~Somiya, H.~Sotani, R.~Takahashi, H.~Takahashi, T.~Akiteru, K.~Taniguchi, A.~Taruya, K.~Tsubono, S.~Tsujikawa, A.~Ueda, K.~i.~Ueda, I.~Watanabe, K.~Yagi, R.~Yamada, S.~Yokoyama, C.~M.~Yoo and Z.~H.~Zhu,
 \tia{Current status of space gravitational wave antenna DECIGO and B-DECIGO} \arX{2006.13545}.
\bibitem{Arzoumanian:2018saf} \au{Z}{Arzoumanian} {et al.} [NANOGRAV], \tia{The NANOGrav 11-year data set: pulsar-timing constraints on the stochastic gravitational-wave background} \doinn{10.3847/1538-4357/aabd3b}{Astrophys.\ J.}{859}{47}{2018} [\arX{1801.02617}].
\bibitem{Janssen:2014dka} \au{G}{Janssen} {et al.}, %, G.~Hobbs, M.~McLaughlin, C.~Bassa, A.~T.~Deller, M.~Kramer, K.~Lee, C.~Mingarelli, P.~Rosado, S.~Sanidas, A.~Sesana, L.~Shao, I.~Stairs, B.~W.~Stappers and J.~Verbiest,
 \tia{Gravitational wave astronomy with the SKA} \doinn{10.22323/1.215.0037}{PoS}{AASKA14}{037}{2015} [\arX{1501.00127}].
\bibitem{KOST1} \au{J}{Khoury}, \au{BA}{Ovrut}, \au{PJ}{Steinhardt}, \au{N}{Turok}, \tia{Ekpyrotic universe: colliding branes and the origin of the hot big bang} \doin{10.1103/PhysRevD.64.123522}{Phys.\ Rev.}{D}{64}{123522}{2001} [\oarX{hep-th/0103239}].
\bibitem{KhSt1} \au{J}{Khoury}, \au{PJ}{Steinhardt}, \tia{Adiabatic ekpyrosis: scale-invariant curvature perturbations from a single scalar field in a contracting universe} \doinn{10.1103/PhysRevLett.104.091301}{Phys.\ Rev.\ Lett.}{104}{091301}{2010} [\arX{0910.2230}].
\bibitem{KhSt2} \au{J}{Khoury}, \au{PJ}{Steinhardt}, \tia{Generating scale-invariant perturbations from rapidly-evolving equation of state} \doin{10.1103/PhysRevD.83.123502}{Phys.\ Rev.}{D}{83}{123502}{2011} [\arX{1101.3548}].
\bibitem{Boyle:2003km} \au{LA}{Boyle}, \au{PJ}{Steinhardt}, \au{N}{Turok}, \tia{Cosmic gravitational-wave background in a cyclic universe} \doin{10.1103/PhysRevD.69.127302}{Phys.\ Rev.}{D}{69}{127302}{2004} [\oarX{hep-th/0307170}].
\bibitem{Kiefer:2011cc} \au{C}{Kiefer}, \au{M}{Kr\"amer}, \tia{Quantum gravitational contributions to the CMB anisotropy spectrum} \doinn{10.1103/PhysRevLett.108.021301}{Phys.\ Rev.\ Lett.}{108}{021301}{2012} [\arX{1103.4967}].
\bibitem{Bini:2013fea} \au{D}{Bini}, \au{G}{Esposito}, \au{C}{Kiefer}, \au{M}{Kr\"amer}, \au{F}{Pessina}, \tia{On the modification of the cosmic microwave background anisotropy spectrum from canonical quantum gravity} \doin{10.1103/PhysRevD.87.104008}{Phys.\ Rev.}{D}{87}{104008}{2013} [\arX{1303.0531}].
\bibitem{Brizuela:2016gnz} \au{D}{Brizuela}, \au{C}{Kiefer}, \au{M}{Kr\"amer}, \tia{Quantum-gravitational effects on gauge-invariant scalar and tensor perturbations during inflation: the slow-roll approximation} \doin{10.1103/PhysRevD.94.123527}{Phys.\ Rev.}{D}{94}{123527}{2016} [\arX{1611.02932}].
\bibitem{Kamenshchik:2017kfs} \au{A.Y}{Kamenshchik}, \au{A}{Tronconi}, \au{G}{Venturi}, \tia{The Born--Oppenheimer method, quantum gravity and matter} \doinn{10.1088/1361-6382/aa8fb3}{Class. Quantum Grav.}{35}{015012}{2018} [\arX{1709.10361}].
\bibitem{Kamenshchik:2015gua} \au{A.Y}{Kamenshchik}, \au{A}{Tronconi}, \au{G}{Venturi}, \tia{Quantum gravity and the large scale anomaly} \doij{10.1088/1475-7516/2015/04/046}{JCAP}{1504}{046}{2015} [\arX{1501.06404}].
\bibitem{Agullo:2015tca} \au{I}{Agull\`o}, \au{NA}{Morris}, \tia{Detailed analysis of the predictions of loop quantum cosmology for the primordial power spectra} \doin{10.1103/PhysRevD.92.124040}{Phys.\ Rev.}{D}{92}{124040}{2015} [\arX{1509.05693}].
\bibitem{Li:2019qzr} \au{BF}{Li}, \au{P}{Singh}, \au{A}{Wang}, \tia{Primordial power spectrum from the dressed metric approach in loop cosmologies} \doin{10.1103/PhysRevD.101.086004}{Phys.\ Rev.}{D}{101}{086004}{2020} [\arX{1912.08225}].
\bibitem{BCT2}  \au{M}{Bojowald}, \au{G}{Calcagni}, \au{S}{Tsujikawa}, \tia{Observational test of inflation in loop quantum cosmology} \doij{10.1088/1475-7516/2011/11/046}{JCAP}{1111}{046}{2011} [\arX{1107.1540}].
\bibitem{Zhu15} \au{T}{Zhu}, \au{A}{Wang}, \au{K}{Kirsten}, \au{G}{Cleaver}, \au{Q}{Sheng}, \au{Q}{Wu}, \tia{Inflationary spectra with inverse-volume corrections in loop quantum cosmology and their observational constraints from Planck 2015 data} \doij{10.1088/1475-7516/2016/03/046}{JCAP}{1603}{046}{2016} [\arX{1510.03855}].
\bibitem{BoBGS} \au{B}{Bolliet}, \au{A}{Barrau}, \au{J}{Grain}, \au{S}{Schander}, \tia{Observational exclusion of a consistent quantum cosmology scenario} \doin{10.1103/PhysRevD.93.124011}{Phys.\ Rev.}{D}{93}{124011}{2016} [\arX{1510.08766}].
\bibitem{deBO}  \au{D}{Mart\'in de Blas}, \au{J}{Olmedo}, \tia{Primordial power spectra for scalar perturbations in loop quantum cosmology} \doij{10.1088/1475-7516/2016/06/029}{JCAP}{06}{029}{2016} [\arX{1601.01716}].
\bibitem{Gomar:2017yww} \au{L}{Castell\'o Gomar}, \au{GA}{Mena Marug\'an}, \au{D}{Mart\'in de Blas}, \au{J}{Olmedo}, \tia{Hybrid loop quantum cosmology and predictions for the cosmic microwave background} \doin{10.1103/PhysRevD.96.103528}{Phys.\ Rev.}{D}{96}{103528}{2017} [\arX{1702.06036}].
\bibitem{Briscese:2013lna}  \au{F}{Briscese}, \au{L}{Modesto}, \au{S}{Tsujikawa}, \tia{Super-renormalizable or finite completion of the Starobinsky theory} \doin{10.1103/PhysRevD.89.024029}{Phys.\ Rev.}{D}{89}{024029}{2014} [\arX{1308.1413}].
\bibitem{Koshelev:2016xqb} \au{AS}{Koshelev}, \au{L}{Modesto}, \au{L}{Rachwa\l}, \au{AA}{Starobinsky}, \tia{Occurrence of exact $R^2$ inflation in non-local UV-complete gravity} \doij{10.1007/JHEP11(2016)067}{JHEP}{11}{067}{2016} [\arX{1604.03127}].
\bibitem{Koshelev:2017tvv}  \au{AS}{Koshelev}, \au{KS}{Kumar}, \au{AA}{Starobinsky}, \tia{$R^2$ inflation to probe non-perturbative quantum gravity} \doij{10.1007/JHEP03(2018)071}{JHEP}{03}{071}{2018} [\arX{1711.08864}].
\bibitem{Koshelev:2020foq} \au{AS}{Koshelev}, \au{KS}{Kumar}, \au{A}{Mazumdar}, \au{AA}{Starobinsky} \tia{Non-Gaussianities and tensor-to-scalar ratio in non-local $R^2$-like inflation} \doij{10.1007/JHEP06(2020)152}{JHEP}{0306}{152}{2020} [\arX{2003.00629}].
\bibitem{CaKu} \au{G}{Calcagni}, \au{S}{Kuroyanagi}, \tia{Stochastic gravitational-wave background in quantum gravity}, \doij{10.1088/1475-7516/2021/03/019}{JCAP}{03}{019}{2021} [\arX{2012.00170}].
\bibitem{tH93}  \au{G}{'t Hooft}, \tia{Dimensional reduction in quantum gravity} \proc{Salamfestschrift}{\au{A}{Ali}, \au{J}{Ellis}, \au{S}{Randjbar-Daemi}}{World Scientific}{Singapore}{1993} [\oarX{gr-qc/9310026}].
\bibitem{Car17} \au{S}{Carlip}, \tia{Dimension and dimensional reduction in quantum gravity} \doinn{10.1088/1361-6382/aa8535}{Class.\ Quant.\ Grav.}{34}{193001}{2017} [\arX{1705.05417}].
\bibitem{ACEMN} \au{G}{Amelino-Camelia}, \au{JR}{Ellis}, \au{NE}{Mavromatos}, \au{DV}{Nanopoulos}, \tia{Distance measurement and wave dispersion in a Liouville string approach to quantum gravity} \doin{10.1142/S0217751X97000566}{Int.\ J.\ Mod.\ Phys.}{A}{12}{607}{1997} [\oarX{hep-th/9605211}].
\bibitem{LaR5}  \au{O}{Lauscher}, \au{M}{Reuter}, \tia{Fractal spacetime structure in asymptotically safe gravity} \doij{10.1088/1126-6708/2005/10/050}{JHEP}{0510}{050}{2005} [\oarX{hep-th/0508202}].
\bibitem{BBL}   \au{A}{Belenchia}, \au{DMT}{Benincasa}, \au{S}{Liberati}, \tia{Nonlocal scalar quantum field theory from causal sets} \doij{10.1007/JHEP03(2015)036}{JHEP}{1503}{036}{2015} [\arX{1411.6513}].
\bibitem{GaPu}  \au{R}{Gambini}, \au{J}{Pullin}, \tia{Nonstandard optics from quantum space-time} \doin{10.1103/PhysRevD.59.124021}{Phys.\ Rev.}{D}{59}{124021}{1999} [\oarX{gr-qc/9809038}].
\bibitem{AMTU}  \au{J}{Alfaro}, \au{HA}{Morales-T\'ecotl}, \au{LF}{Urrutia}, \tia{Quantum gravity corrections to neutrino propagation} \doinn{10.1103/PhysRevLett.84.2318}{Phys.\ Rev.\ Lett.}{84}{2318}{2000} [\oarX{gr-qc/9909079}].
\bibitem{ACAP}  \au{G}{Amelino-Camelia}, \au{M}{Arzano}, \au{A}{Procaccini}, \tia{Severe constraints on loop-quantum-gravity energy-momentum dispersion relation from black-hole area-entropy law} \doin{10.1103/PhysRevD.70.107501}{Phys.\ Rev.}{D}{70}{107501}{2004} [\oarX{gr-qc/0405084}].
\bibitem{Ron16} \au{M}{Ronco}, \tia{On the UV dimensions of loop quantum gravity} \doinn{10.1155/2016/9897051}{Adv.\ High Energy Phys.}{2016}{9897051}{2016} [\arX{1605.05979}].
\bibitem{CDL} \au{V}{Cardoso}, \au{\'OJC}{Dias}, \au{JPS}{Lemos}, \tia{Gravitational radiation in $D$-dimensional spacetimes} \doin{10.1103/PhysRevD.67.064026}{Phys.\ Rev.}{D}{67}{064026}{2003} [\oarX{hep-th/0212168}].
\bibitem{Mag07} \au{M}{Maggiore}, \book{Gravitational Waves, Vol.\ 1}{Oxford University Press}{Oxford}{UK}{2007}.
\bibitem{NgDa}  \au{YJ}{Ng}, \au{H}{Van Dam}, \tia{Limit to space-time measurement} \doin{10.1142/S0217732394000356}{Mod.\ Phys.\ Lett.}{A}{9}{335}{1994}.
\bibitem{Ame94} \au{G}{Amelino-Camelia}, \tia{Limits on the measurability of space-time distances in the semiclassical approximation of quantum gravity} \doin{10.1142/S0217732394003245 }{Mod.\ Phys.\ Lett.}{A}{9}{3415}{1994} [\arX{gr-qc/9603014}].
%\bibitem{first} \au{G}{Calcagni}, \tia{Multiscale spacetimes from first principles} \doin{10.1103/PhysRevD.95.064057}{Phys.\ Rev.}{D}{95}{064057}{2017} [\arX{1609.02776}].
\bibitem{DeMe}  \au{C}{Deffayet}, \au{K}{Menou}, \tia{Probing gravity with spacetime sirens} \doinn{10.1086/522931}{Astrophys.\ J.}{668}{L143}{2007} [\arX{0709.0003}].
\bibitem{PFHS}  \au{K}{Pardo}, \au{M}{Fishbach}, \au{DE}{Holz}, \au{DN}{Spergel}, \tia{Limits on the number of spacetime dimensions from GW170817} \doij{10.1088/1475-7516/2018/07/048}{JCAP}{1807}{048}{2018} [\arX{1801.08160}].
\bibitem{Andriot:2017oaz} \au{D}{Andriot}, \au{G}{Lucena G\'omez}, \tia{Signatures of extra dimensions in gravitational waves} \doij{10.1088/1475-7516/2017/06/048}{JCAP}{1706}{048}{2017} [\arX{1704.07392}].
\bibitem{Abb18} \au{BP}{Abbott} {et al.} [LIGO Scientific and Virgo Collaborations], \tia{Tests of general relativity with GW170817} \doinn{10.1103/PhysRevLett.123.011102}{Phys.\ Rev.\ Lett.}{123}{011102}{2019} [\arX{1811.00364}].
\bibitem{Ab17b} \au{BP}{Abbott} {et al.} [LIGO Scientific and Virgo and  Fermi-GBM and INTEGRAL Collaborations], \tia{Gravitational waves and gamma-rays from a binary neutron star merger: GW170817 and GRB 170817A} \doinn{10.3847/2041-8213/aa920c}{Astrophys.\ J.}{848}{L13}{2017} [\arX{1710.05834}].
\bibitem{COT3} \au{G}{Calcagni}, \au{D}{Oriti}, \au{J}{Th\"urigen}, \tia{Dimensional flow in discrete quantum geometries} \doin{10.1103/PhysRevD.91.084047}{Phys.\ Rev.}{D}{91}{084047}{2015} [\arX{1412.8390}].
\bibitem{Dalal:2006qt}  \au{N}{Dalal}, \au{DE}{Holz}, \au{SA}{Hughes}, \au{B}{Jain}, \tia{Short GRB and binary black hole standard sirens as a probe of dark energy} \doin{10.1103/PhysRevD.74.063006}{Phys.\ Rev.}{D}{74}{063006}{2006} [\oarX{astro-ph/0601275}].
\bibitem{Nissanke:2009kt} \au{S}{Nissanke}, \au{DE}{Holz}, \au{SA}{Hughes}, \au{N}{Dalal}, \au{JL}{Sievers}, \tia{Exploring short gamma-ray bursts as gravitational-wave standard sirens} \doinn{10.1088/0004-637X/725/1/496}{Astrophys.\ J.}{725}{496}{2010} [\arX{0904.1017}].
\bibitem{Camera:2013xfa}  \au{S}{Camera}, \au{A}{Nishizawa}, \tia{Beyond concordance cosmology with magnification of gravitational-wave standard sirens} \doinn{10.1103/PhysRevLett.110.151103}{Phys.\ Rev.\ Lett.}{110}{151103}{2013} [\arX{1303.5446}].
\bibitem{Tamanini:2016zlh} \au{N}{Tamanini}, \au{C}{Caprini}, \au{E}{Barausse}, \au{A}{Sesana}, \au{A}{Klein}, \au{A}{Petiteau}, \tia{Science with the space-based interferometer eLISA. III: Probing the expansion of the universe using gravitational wave standard sirens} \doij{10.1088/1475-7516/2016/04/002}{JCAP}{1604}{002}{2016} [\arX{1601.07112}].
\bibitem{Gasperini:2016gre} \au{M}{Gasperini}, \tia{Observable gravitational waves in pre-big bang cosmology: an update} \doij{10.1088/1475-7516/2016/12/010}{JCAP}{1612}{010}{2016} [\arX{1606.07889}].
%\bibitem{basic-contrib} Brown B, Aaron M (2001) The politics of nature. In: Smith J (ed) The rise of modern genomics, 3rd edn. Wiley, New York, p 234--295  
%\bibitem{basic-online} Dod J (1999) Effective Substances. In: The dictionary of substances and their effects. Royal Society of Chemistry. Available via DIALOG. \\
%\url{http://www.rsc.org/dose/title of subordinate document. Cited 15 Jan 1999}
%\bibitem{basic-DOI} Slifka MK, Whitton JL (2000) Clinical implications of dysregulated cytokine production. J Mol Med, doi: 10.1007/s001090000086
%\bibitem{basic-journal} Smith J, Jones M Jr, Houghton L et al (1999) Future of health insurance. N Engl J Med 965:325--329
%\bibitem{basic-mono} South J, Blass B (2001) The future of modern genomics. Blackwell, London 
\end{thebibliography}
%

\end{document}